\documentclass[a4paper,11pt]{article}
\usepackage[utf8]{inputenc}
\usepackage{jheppub, bm, color}
\usepackage{amssymb, amsfonts, slashed, amsthm, amsmath, graphicx, soul, empheq, cancel, physics, caption, subcaption}
\usepackage{listings}
\usepackage{xcolor}
\preprint{{\color{white}---}}

\title{
 Exploring the Landscape of Spontaneous CP Violation \\ in Supersymmetric Theories
}

\author{Fangchao Liu,  Shota Nakagawa, Yuichiro Nakai, and Yaoduo Wang}
\affiliation{Tsung-Dao Lee Institute, Shanghai Jiao Tong University, \\
No.~1 Lisuo Road, Pudong New Area, Shanghai 201210, China}
\affiliation{School of Physics and Astronomy, Shanghai Jiao Tong University, \\
800 Dongchuan Road, Shanghai 200240, China}

\emailAdd{liufangchao@sjtu.edu.cn}
\emailAdd{shota.nakagawa@sjtu.edu.cn}
\emailAdd{ynakai@sjtu.edu.cn}
\emailAdd{yaoduowang@sjtu.edu.cn}

\abstract{The strong CP problem remains one of the most important unresolved issues in the Standard Model. Spontaneous CP violation (SCPV) is a promising approach to the problem by assuming that CP is an exact symmetry of the Lagrangian but broken spontaneously at the vacuum, which enables the generation of the observed Cabibbo-Kobayashi-Maskawa (CKM) phase without reintroducing a nonzero strong CP phase. Supersymmetry (SUSY) provides a natural framework to accommodate such a mechanism, as SUSY can not only protect the scale of SCPV from radiative corrections but also suppress problematic higher-dimensional operators generating a strong CP phase.
In the present study, we explore the realization of SCPV in two distinct SUSY scenarios. First, we investigate SCPV in the exact SUSY limit by extending the spurion formalism developed in non-supersymmetric theories to identify the necessary condition for stabilizing CP-violating phases, and by analyzing the stabilization of radial vacuum expectation values through R-symmetry constraints on the superpotential. Second, we construct a model in which CP is spontaneously broken at an intermediate scale along pseudo-flat directions, stabilized by soft SUSY breaking and non-perturbative effects of a gauge theory. The latter setup predicts light scalars in the SCPV sector whose masses are determined by the SUSY breaking scale.}

\begin{document}

\newcommand{\eV}{ \ {\rm eV} }
\newcommand{\KeV}{ \ {\rm keV} }
\newcommand{\MeV}{\  {\rm MeV} }
\newcommand{\GeV}{\  {\rm GeV} }
\newcommand{\TeV}{\  {\rm TeV} }
\newcommand{\1}{\mbox{1}\hspace{-0.25em}\mbox{l}}
\newcommand{\Red}[1]{{\color{red} {#1}}}

\newcommand{\lmk}{\left(}  
\newcommand{\rmk}{\right)}
\newcommand{\lkk}{\left[}  
\newcommand{\rkk}{\right]}
\newcommand{\lhk}{\left \{ }  
\newcommand{\rhk}{\right \} }
\newcommand{\del}{\partial}  
\newcommand{\la}{\left\langle} 
\newcommand{\ra}{\right\rangle}
\newcommand{\half}{\frac{1}{2}}

\newcommand{\bea}{\begin{array}}
\newcommand{\eea}{\end{array}}
\newcommand{\beq}{\begin{eqnarray}}
\newcommand{\eeq}{\end{eqnarray}}
\newcommand{\eq}[1]{Eq.~(\ref{#1})}

\newcommand{\Mpl}{M_{\rm Pl}}
\newcommand{\mg}{m_{3/2}}
\newcommand{\mphi}{m_{\phi}}
\newcommand{\Hz}{\ {\rm Hz}}
\newcommand{\for}{\quad \text{for }}
\newcommand{\Min}{\text{Min}}
\newcommand{\Max}{\text{Max}}
\newcommand{\Kahler}{K\"{a}hler }
\newcommand{\cphi}{\varphi}
\newcommand{\diag}{{\rm diag}}

\newcommand{\SUf}{SU(3)_{\rm f}}
\newcommand{\Upq}{U(1)_{\rm PQ}}
\newcommand{\Zpq}{Z^{\rm PQ}_3}
\newcommand{\Cpq}{C_{\rm PQ}}
\newcommand{\ubar}{u^c}
\newcommand{\dbar}{d^c}
\newcommand{\ebar}{e^c}
\newcommand{\nubar}{\nu^c}
\newcommand{\Ndw}{N_{\rm DW}}
\newcommand{\Fpq}{F_{\rm PQ}}
\newcommand{\fpq}{v_{\rm PQ}}
\newcommand{\Br}{{\rm Br}}
\newcommand{\Lag}{\mathcal{L}}
\newcommand{\Lqcd}{\Lambda_{\rm QCD}}

\newcommand{\ji}{j_{\rm inf}} 
\newcommand{\jb}{j_{B-L}} 
\newcommand{\M}{M} 
\newcommand{\im}{{\rm Im} }
\newcommand{\re}{{\rm Re} }

\def\lrf#1#2{ \left(\frac{#1}{#2}\right)}
\def\lrfp#1#2#3{ \left(\frac{#1}{#2} \right)^{#3}}
\def\lrp#1#2{\left( #1 \right)^{#2}}
\def\REF#1{Ref.~\cite{#1}}
\def\SEC#1{Sec.~\ref{#1}}
\def\FIG#1{Fig.~\ref{#1}}
\def\EQ#1{Eq.~(\ref{#1})}
\def\EQS#1{Eqs.~(\ref{#1})}
\def\TEV#1{10^{#1}{\rm\,TeV}}
\def\GEV#1{10^{#1}{\rm\,GeV}}
\def\MEV#1{10^{#1}{\rm\,MeV}}
\def\KEV#1{10^{#1}{\rm\,keV}}
\def\blue#1{\textcolor{blue}{#1}}
\def\red#1{\textcolor{blue}{#1}}

\newcommand{\eff}{\Delta N_{\rm eff}}
\newcommand{\neff}{\Delta N_{\rm eff}}
\newcommand{\cc}{\Omega_\Lambda}
\newcommand{\Mpc}{\ {\rm Mpc}}
\newcommand{\Msolar}{M_\odot}

\def\sn#1{\textcolor{red}{#1}}
\def\SN#1{\textcolor{red}{[{\bf SN:} #1]}}
\def\FL#1{\textcolor{blue}{[{\bf FL:} #1]}}

\maketitle
\flushbottom

\section{Introduction}\label{intro}

The strong CP problem remains one of the most profound puzzles in the Standard Model (SM). The quantum chromodynamics (QCD) Lagrangian admits a CP‑violating parameter $\bar{\theta}$, defined as the sum of the Yang-Mills vacuum angle and the complex phase of the determinant of the quark Yukawa matrices, while experimental limits on the neutron electric dipole moment (EDM) require $|\bar{\theta}|\lesssim 10^{-10}$
\cite{Baker:2006ts,Pendlebury:2015lrz,Abel:2020pzs}, 
an extraordinarily small value that lacks an intrinsic explanation within the SM. 
An appealing approach to the problem assumes that CP is an exact symmetry of the fundamental Lagrangian and that the observed CP-violating Cabibbo-Kobayashi-Maskawa (CKM) phase originates from complex vacuum expectation values (VEVs) of scalar fields which spontaneously break the CP symmetry.
A well-known explicit realization of this idea is the Nelson-Barr mechanism~\cite{Nelson:1983zb,Barr:1984qx,Barr:1984fh}, in which a vector-like pair of heavy quarks is added to the SM
\cite{Bento:1991ez}, and the extended quark mass matrix transmits spontaneous CP violation (SCPV) into the CKM matrix without reintroducing $\bar{\theta}$. 
However, this mechanism typically requires new scalar fields whose VEVs break CP at a scale hierarchically below the Planck scale. 
It also suffers from sensitivity to higher-dimensional operators and radiative corrections that can regenerate a strong CP phase, thereby undermining the solution
\cite{Dine:2015jga}. 
Supersymmetry (SUSY) offers a natural framework to address these difficulties
\cite{Barr:1993hb,Dine:1993qm,Evans:2020vil,Fujikura:2022sot,Feruglio_2023,Feruglio:2024ytl,Feruglio:2024dnc,Feruglio:2025ajb}.\footnote{
For non-supersymmetric models to address the issues in the Nelson-Barr mechanism, see e.g. Refs.~\cite{Vecchi:2014hpa,Valenti:2021xjp,Girmohanta:2022giy,Asadi:2022vys,Bai:2022nat,Murai:2024alz,Murai:2024bjy,Ferro-Hernandez:2024snl,Jiang:2024frx}.
} 
By protecting scalar masses from large radiative corrections, SUSY stabilizes the scale of SCPV in much the same way it stabilizes the electroweak scale. 
In addition, the holomorphy of the superpotential and the SUSY non-renormalization theorem can forbid or strongly suppress dangerous higher-dimensional operators. 
Besides the Nelson-Barr scenario, the Hiller-Schmaltz mechanism~\cite{Hiller:2001qg,Hiller:2002um} exploits a special property of SUSY: while the Kähler potential is renormalized, the hermiticity of wave-function renormalization factors protects the strong CP phase from loop corrections even as the CKM phase receives nonzero contributions.
The observed CKM phase requires a large Yukawa coupling, but this problem as well as the scalegenesis of SCPV can be dynamically addressed in a SUSY QCD model \cite{Nakagawa:2024ddd}.

A distinctive feature of supersymmetric theories is the generic presence of flat directions, valleys in field space along which the scalar potential is exactly flat (even under perturbative quantum corrections) in the SUSY limit. 
Such flat directions naturally contain a minimum where complex VEVs of the scalar fields spontaneously break CP symmetry. 
The potential minimum has to be stabilized, lifting up the flat directions with positive masses for all scalar fields around the minimum,
but there are two qualitatively distinct ways for the stabilization:
either purely by supersymmetric dynamics or through SUSY-breaking effects
\cite{Dine:2015jga}. 
The former scenario can be realized by introducing proper superpotential terms (see e.g. Refs.~\cite{Dine:2015jga,Nakagawa:2024ddd}).
However, it is still unclear what kind of superpotential terms can stabilize a vacuum with SCPV in general, both in the phase directions responsible for CP violation and in the radial directions that determine the existence of isolated vacua.
For non-supersymmetric theories, the authors of Ref.~\cite{Haber:2012np} developed a spurion formalism to find the necessary condition for SCPV that requires a sufficient number of inequivalent spurion fields of $U(1)$ symmetries associated with involved scalar fields. For instance, a single complex scalar field necessitates at least two spurions with different $U(1)$ charges to stabilize its complex phase at a nonzero value.
We then aim to extend this formalism to supersymmetric settings and find a condition that the superpotential is required to satisfy for the supersymmetric realization of SCPV. In particular, while the spurion analysis constrains the stabilization of CP-violating phases, we also analyze the role of R-symmetry in controlling the stabilization of radial vacuum expectation values.
To this end, in the present paper, we develop a systematic operator formalism that maps spurions in the superpotential to the scalar potential counterparts, and combine it with an R-charge analysis of the superpotential, enabling the identification and systematic construction of supersymmetric models that realize SCPV with given symmetries by extending the argument of Ref.~\cite{Haber:2012np}.

For the supersymmetric realization of SCPV, all fields in the SCPV sector have masses naturally
at an intermediate scale of SCPV which is generally much higher than the electroweak scale,
so that its experimental probe is difficult.
In contrast, if some flat directions around a minimum with SCPV are only lifted up by SUSY breaking effects,
such a SCPV sector contains fields whose masses are determined by the size of SUSY breaking typically much smaller than
the scale of SCPV.
Since the SCPV sector must communicate with the visible SM sector to transmit SCPV into the CKM matrix
as in the Nelson-Barr mechanism, these light states should interact with the SM fields,
potentially offering experimental signatures.
In the present work, we also pursue this exciting scenario 
and construct a concrete model realizing SCPV through SUSY breaking effects for the first time to the best of our knowledge.

The rest of the present paper is organized as follows.
Section~\ref{sec:SUSYconserv} analyzes SCPV in the exact SUSY limit by extending the spurion formalism developed in non-supersymmetric theories and by incorporating an R-charge analysis of the superpotential, establishing criteria to determine whether a given superpotential satisfies the necessary conditions for SCPV. In section~\ref{susy-breaking case}, we construct a model in which CP symmetry is spontaneously broken at an intermediate scale along pseudo-flat directions. The vacuum is stabilized by combined effects of soft SUSY breaking and non-perturbative dynamics of a gauge theory, leading to light scalar modes in the SCPV sector whose masses are set by the soft SUSY-breaking scale.
Section~\ref{conclusion} is devoted to conclusions and discussions.
Some details and program codes are summarized in appendices.

\section{Supersymmetric realization of SCPV
\label{sec:SUSYconserv}}

In supersymmetric theories, the vacuum structure is governed by the form of the superpotential. To identify when CP-violating vacua can arise, we develop a systematic procedure based on a spurion formalism \cite{Haber:2012np} applied directly to the superpotential, which characterizes the explicit breaking of phase redefinition symmetries and provides the necessary conditions for stabilizing CP-violating phases.
The stabilization of the radial direction  is instead constrained by the requirement of supersymmetric vacua, namely, the existence of solutions to the F-term conditions with unbroken SUSY. We analyze this condition using R-charge assignments that require the vanishing VEV of R-charged fields \cite{Nelson:1993nf}. Together, these two independent criteria give the necessary condition for the existence of physically viable CP-violating supersymmetric vacua.

\subsection{Spurion analysis: the non-supersymmetric case}\label{non-susy}
We start the discussion by briefly reviewing the non-supersymmetric case \cite{Haber:2012np}, where one or more complex scalars serve as the source of CP violation. 
A Lagrangian is said to be on a real basis when all its parameters are set to be real, and thus the CP symmetry is preserved at the Lagrangian level.
Given a scalar potential on a real basis, SCPV is then triggered by nonzero phases among VEV(s) of complex scalar(s).
However, physical CP-violating phases must immune to any field redefinition, which may not be obvious especially for multi-field cases.

The single complex scalar case would be a good example to demonstrate the essence.
Consider a complex scalar field $\phi$. This
is associated with a field redefinition $\phi \to e^{i\theta}\phi$,
which can be parameterized by a subgroup $\mathcal{H} = U(1)_\phi$ of the maximal global $O(2)$ symmetry in the presence of the kinetic term.
Under this field redefinition subgroup $\mathcal{H}$, $\phi$ is charged by unity.
When $\phi$ obtains a VEV $\langle \phi \rangle$ in a scalar potential $V(\phi)$,
the phase of $\langle \phi \rangle$ is physical and non-vanishing if and only if $V(\phi)$ satisfies the following two requirements:
\begin{itemize}
    \item The field redefinition subgroup $\mathcal{H}$ is explicitly broken by $V(\phi)$.
    \item Up to a discrete transformation that flips signs of parameters, the phase of  $\langle \phi \rangle$ is not stabilized somewhere equivalent to zero or $\pi$.
\end{itemize}
The first requirement ensures that the phase of  $\langle \phi \rangle$ cannot be rotated away by the field redefinition.
The second requirement guarantees that CP is spontaneously broken.
Now we examine these two requirements with some simplest forms of $V(\phi)$ and show the general patterns.
As a first attempt to meet the first requirement, suppose $\mathcal{H}$ is explicitly broken by a single quadratic term,
\beq 
V(\phi) = V_0(|\phi|) +  \tilde s_2 \phi^2 + \rm{h.c.},
  \label{scalar-pot1}
\eeq 
where $\tilde s_2$ is a real parameter.
By parameterizing $\phi=v e^{i\theta}$, the scalar potential becomes
\beq 
V(\phi) = V_0(v) + 2 \tilde s_2 v^2 \cos(2\theta)  .
  \label{scalar-pot1paramed}
\eeq 
It turns out that the phase of $\langle \phi \rangle$ stabilizes at $\theta = \pi/2$ ($\theta=0$) for $\tilde s_2>0$ ($\tilde s_2<0$).
Unfortunately, this means that the second requirement is not satisfied by the scalar potential~\eqref{scalar-pot1}, as $\theta=\pi/2$ is equivalent to $\theta=0$ up to a discrete transformation $\tilde s_2 \to -\tilde s_2, \phi \to  \pm i \phi$.
The solution is to include an extra term charged ``inequivalently" under $\mathcal{H}$, e.g. a quartic term,
\beq 
V(\phi) = V_0(|\phi|) +  \tilde s_2 \phi^2 + \tilde s_4 \phi^4 + \rm{h.c.} ,
  \label{scalar-pot2}
\eeq 
where both $\tilde s_2$ and $\tilde s_4$ are real valued.
Again, we write the scalar potential in terms of the parameterization  $\phi=v e^{i\theta}$,
\beq 
V(\phi) = V_0(v) + 2 \tilde s_2 v^2 \cos2\theta + 2 \tilde s_4 v^4 \cos4\theta .
  \label{scalar-pot2paramed}
\eeq 
In this case, the phase of $\langle \phi \rangle$ is stabilized at $\theta = \frac12 \arccos\frac{-\tilde s_2}{4 \tilde s_4 v^2}$ for $|\tilde s_2 / \tilde s_4 |<4v^2$.
To verify that the second requirement is indeed satisfied, we write down the only allowed sign-flipping discrete transformation on Eq.~\eqref{scalar-pot2} explicitly as $\tilde s_2\to -\tilde s_2, \ \tilde s_4\to \tilde s_4, \  \phi\to \pm i\phi$.
This transformation can shift  $\theta = \frac12 \arccos\frac{-\tilde s_2}{4 \tilde s_4 v^2}$ to neither zero nor $\pi$.
Therefore we conclude that the scalar potential~\eqref{scalar-pot2} produces a physical CP-violating phase.
In summary, to obtain a physical CP phase, we need the incorporation of two types of terms in a scalar potential:  one is responsible to break $\mathcal{H}$ explicitly, and the other plays the role to support the CP phase from being shifted to zero or $\pi$.
In other words, a scalar potential that can spontaneously break CP is of the standard form,
\beq 
V(\phi) = V_0(|\phi|) + V_{\text{break}}(\phi) + V_{\text{support}}(\phi) \ .
\label{standard-pot}
\eeq 
One can take $V_{\text{break}}(\phi)=\tilde s_2 \phi^2 + \rm{h.c.}$ and $ V_{\text{support}}(\phi)=\tilde s_4 \phi^4 + \rm{h.c.}$ to reproduce \EQ{scalar-pot2}.
Note that there is an arbitration to choose $V_{\text{break}}(\phi)$, for example it is
also possible to take $V_{\text{break}}(\phi)=\tilde s_4 \phi^4 + \rm{h.c.}$ and leave the others to $V_{\text{break}}(\phi)$.
There is no status difference for terms in $V_{\text{break}}(\phi)$ and $V_{\text{support}}(\phi)$ but the fact that they break the field redefinition subgroup $\mathcal{H}$ in ``inequivalent" ways matters.

To make the concept of the ``inequivalency" more rigorous and generalize the statement to multi-scalar cases, it is useful to utilize the spurion technique.
The idea of spurions is to study explicitly broken symmetries as if they are unbroken.
This can be done by artificially charging the coefficients of symmetry breaking terms under the broken symmetries.
These coefficients are then denoted as spurions.
One can also reinterpret these spurions as VEVs that spontaneously break the symmetries.
In our previous example of \EQ{scalar-pot2}, parameters $\tilde s_2$ and $\tilde s_4$ (and their hermitian conjugates) play the role of spurions and carry charges of magnitude $2$ and $4$ under $\mathcal{H}$ respectively.
Here we clarify the definition of ``inequivalency":
two spurions are equivalent if and only if they carry the same charge under $\mathcal{H}$ up to an overall sign.
Since $\tilde s_2$ and $\tilde s_4$ carry non-vanishing $\mathcal{H}$ charges and are not equal to each other up to an overall sign,
they are inequivalent and able to produce a physical CP phase.
For a scalar potential $V(\phi)$, a general observation is that at least two inequivalent spurions are needed to guarantee that CP is spontaneously broken.

Now we are ready to dive into multi-scalar cases.
In the case of $N$ complex scalars $\phi_1, \phi_2, \cdots, \phi_N$, the maximal symmetry group in the presence of the kinetic term is $O(2N)$.
The independent phases of the scalars can be parameterized by a field redefinition subgroup $\mathcal{H} = U(1)_1 \times U(1)_2 \times \cdots \times U(1)_N$, where the $U(1)_i$ rotates the phase of the $i$-th scalar $\phi_i$.
A general scalar potential can be written as
\beq 
V(\phi) = V_0 \left(|\phi| \right) + \sum_{l=1}^{N_s}\sum_{m} \tilde s^{\bm{Q}_{(l)}}_m  \mathcal{\tilde F}_{\bm{Q}_{(l)}}^m \left(\phi \right),
\label{scalar-pot-multi}
\eeq
where $\tilde  s^{\bm{Q}_{(l)}}_m$ denotes the spurion in the $l$-th {inequivalent} spurion class that carries the charge $\bm{Q}_{(l)}$ under $\mathcal{H}$ of the $m$-th multiplicity and $\mathcal{\tilde F}_{\bm{Q}_{(l)}}^m \left(\phi \right)$ is the monomial of scalars associated with the spurion $\tilde s^{\bm{Q}_{(l)}}_m$.
We can determine the number of physical CP-violating phases by rewriting the multi-scalar potential  $V(\phi)$ into the standard form analogously to \EQ{standard-pot},
\beq 
 V(\phi ) = V_0 \left(|\phi| \right) + V_{\text{break}}\left(\phi \right) + V_{\text{support}}\left(\phi \right),
 \label{standard-pot-multi}
 \eeq 
 where $\phi$ represents the collection of $N$ scalars.
To do so, it is helpful to collect all inequivalent spurion classes in $V(\phi)$  into a $N_s\times N$ charge matrix $\mathcal Q$ whose elements are given by the charges of spurions under each of $U(1)_i$'s:

\beq 
\mathcal Q =
\begin{pmatrix}
\bm{Q}_{(1)1} & \bm{Q}_{(1)2} & \cdots & \bm{Q}_{(1)N}\\
\bm{Q}_{(2)1} & \bm{Q}_{(2)2} & \cdots & \bm{Q}_{(2)N}\\
\vdots & \vdots & \ddots & \vdots \\
\bm{Q}_{(N_s)1} & \bm{Q}_{(N_s)2} & \cdots & \bm{Q}_{(N_s)N}
\end{pmatrix}.
\eeq 

It is possible to diagonalize $\mathcal{Q}$ through a redefinition of $\mathcal{H}$ as $ U(1)_1 \times U(1)_2 \times \cdots \times U(1)_N$ $\to  U(1)'_1 \times U(1)'_2 \times \cdots U(1)'_N$. 
Given the rank of the charge matrix, ${\rm{rank}} \, \mathcal{Q}=r$,
this is equivalent to rewrite $\mathcal{Q}$ into the standard form by column transformations,
\beq \label{Q'}
\mathcal Q' = \mathcal{Q} \mathcal Q^{-1} = 
\begin{pmatrix}
\mathcal{Q}_{r\times r}^{\text{break}} & \bm{0}_{r\times (N-r)}\\
 \mathcal Q_{(N_s-r)\times r}^{\text{support}} & \bm{0}_{(N_s-r)\times (N-r)}
\end{pmatrix},
\eeq 
where $\mathcal Q^{-1}$ is the right pseudo-inverse of $\mathcal{Q}$. 
The diagonal part of $\mathcal{Q}'$ is denoted as $\mathcal{Q}^\text{break} = \bm 1$ and the remanent part   is referred to as the supporting matrix $\mathcal{Q}^\text{support}$ in the following discussion.
Again, we parameterize the scalars $\phi_k = v_k e^{i X'_j \theta_j}$, where $X_j' \equiv X_i \mathcal{Q}^{-1}_{ij}$ is the generator of $U(1)'_j$ in $\mathcal{H}$.
Then it is straightforward to show that $\mathcal{Q}^\text{break} \, (= \bm{1})$ and $\mathcal{Q}^\text{support}$ contribute to $V_{\text{break}}\left(\phi \right)$ and $V_{\text{support}}\left(\phi \right)$ as
\begin{equation}
\begin{split}
&V_{\text{break}}\left(\phi \right) = \sum_{l=1}^{r}\sum_{m} \tilde s^{\bm{Q}_{(l)}}_m   \mathcal{\tilde F}_{\bm{Q}_{(l)}}^m \left(v\right)  \cos\theta_l \ , \\[1ex]
&V_{\text{support}}\left(\phi \right) = \sum_{l=r+1}^{N_s}\sum_{m} \tilde s^{\bm{Q}_{(l)}}_m  \mathcal{\tilde F}_{\bm{Q}_{(l)}}^m \left(v\right)  \cos \sum_{j} \mathcal{Q}^\text{support}_{(l-r), \, j} \theta_j \ ,
\end{split}
\end{equation}
respectively.
Following the discussion of the single scalar case, the potential CP-violating phases are just those phases that explicitly appear in both  $V_{\text{break}}\left(\phi \right)$ and $V_{\text{support}}\left(\phi \right)$.
In conclusion, the number of CP-violating phases $d$ obtained from the general scalar potential~\eqref{scalar-pot-multi} can be found by counting the number of $U(1)_i$'s that are explicitly broken and supported by $\mathcal{Q}^\text{support}$,
\beq 
d = \sum_{l=1}^r \Theta \left( \left|\mathcal{Q}^\text{support}_{l} \right| \right),
\label{nCPphase}
\eeq 
where $|\mathcal{Q}^\text{support}_{l}|$ is the norm of the $l$-th column of the supporting matrix $\mathcal{Q}^\text{support}$ and $\Theta$ is the Heaviside step function,
\beq 
\Theta(x) = 
\begin{cases}
    1, \quad x>0\\
    0, \quad x\le 0
\end{cases}.
\eeq

\subsection{Spurion analysis:
the supersymmetric case}\label{sec:susycase}

Having reviewed the non-supersymmetric case, we now formalize a general superpotential $W$ of $N$ chiral superfields $\Phi_{i=1,2,\cdots, N}$ in terms of spurions,
\begin{align}
  W =  \sum_{\bm{Q}  \in \mathbb{N}^N}  s^{\bm{Q}} \mathcal{F}_{\bm{Q}} (\Phi) \ ,
  \label{Wexpansion}
\end{align}
where $s^{\bm{Q}}$ denotes the spurion that carries the charge $\bm Q$ with $\bm Q \in \mathbb{N}^N$ being a general charge vector under the redefinition subgroup of the chiral superfields, $\mathcal{H} = U(1)_1\times U(1)_2 \times \cdots \times U(1)_N$.
Here $\mathbb N = \{0,1,2\cdots\}$ is the set of natural numbers.
And $\mathcal{F}_{\bm{Q}}(\Phi)$ is the monomial of the chiral superfields $\Phi_{i=1,2,\cdots, N}$ associated with the spurion $s^{\bm Q}$.
As an example, taking $N=1$ and setting all $s^{\bm{Q}}=0$ except $s^1=L$, $s^2=M/2$ and $s^3=Y/6$, one obtains a renormalizable superpotential of a single chiral superfield, $W=L\Phi + \frac12 M \Phi^2 + \frac16 Y \Phi^3$.
The multi-field case is shown independently in  Appendix \ref{sec:spurion}.
Note that there is no multiplicity\footnote{For multi-field cases, spurions whose indices are equal up to a permutation are regarded as identical, e.g. $M_{12}$ and $M_{21}$ are referred to as the same spurion for $W\supset M_{ij} \Phi_i \Phi_j$ for $i,j=1,2$. }  for spurions and monomials in any superpotential $W$ since $W$ is holomorphic.
After integrating out the auxiliary field $F$, one can extract the scalar potential from \EQ{Wexpansion},
\begin{equation}\
\label{mapW2V1}
    V(\phi, \phi^{*}) 
=\sum_i \left|\frac{\del W}{\del\phi_i}\right|^2.
\end{equation}
Here $\phi_i$ and $\phi^*_i$ are the scalar component of $\Phi_i$ and its complex conjugation, respectively.
The conclusion of \EQ{nCPphase} is still applicable to the scalar potential~\eqref{mapW2V1}, 
as long as the spurions $s^{\bm{Q}}$ in the superpotential $W$ are properly mapped into the spurions in the scalar potential $V$ as in \EQ{scalar-pot-multi}.
This can be done by using the mapping relation between superpotential spurions $s^{\bm Q}$ and scalar potential spurions $\tilde s^{\bm Q}$, 
\begin{equation}\label{Vspurion}
\tilde{s}^{\bm{Q}}_{\bm{Q}'} = \sum_{i=1}^{N} \tilde{s}^{\bm{Q}}_{i,\bm{Q}'} = 
\sum_{i=1}^{N} 
s^{\bm{Q}'+\bm{q}_i} (s^{\bm{Q}'-\bm{Q}+\bm{q}_i})^*.
\end{equation}
The \EQ{Vspurion} shows clear correspondence to \EQ{mapW2V1} and we will prove it in the next subsection.
The superpotential spurion $s^{\bm Q'+\bm q_i}$ corresponds to charge $\bm Q'$ shifted by a unit charge vector $\bm q_i=(0,...,0,\underset{i}{1},0,...,0)$ of $U(1)_i$.
This can be understood as the differential operation $\frac{\partial}{\partial \phi_i}$ carries the charge $-\bm q_i$.
The hermitian conjugation of $ s^{\bm{Q}'-\bm{Q}+\bm{q}_i}$ carries the charge of $-\bm{Q}'+\bm{Q}-\bm{q}_i$, which guarantees  that the scalar potential spurion $\tilde s^{\bm Q}$ is properly charged.
In \EQ{Vspurion}, multiple scalar potential spurions carrying the same charge, i.e. multiplicities of inequivalent spurion classes are distinguished by  ${\bm Q}' \in \mathbb{Z}^N$.
By using the relation~\eqref{Vspurion}, one can rewrite the scalar potential $V$ in terms of scalar potential spurions $\tilde s$,
\begin{equation}
    V(\phi, \phi^*) = \sum_{\bm{Q,Q}'}  \tilde{s}^{\bm{Q}}_{\bm{Q}'} \mathcal{\tilde{F}}_{{\bm{Q}'} }(\phi) \mathcal{\tilde{F}}_{\bm{Q}'-\bm{Q}}^*(\phi) \ ,\label{mapW2V2}
\end{equation}
where $\mathcal{\tilde{F}}_{\bm{Q}'-\bm{Q}}^*(\phi)$ is the complex conjugation of $\mathcal{\tilde{F}}_{\bm{Q}'-\bm{Q}}(\phi)$.
Compared with direct calculation of \EQ{mapW2V1}, \EQ{Vspurion} can be more practical as multiplicities of inequivalent spurion classes are sorted automatically so that it can be straightforward to write down the general charge matrix for any given superpotential $W$,
\begin{equation}
    \mathcal{Q}_{(l)}= \bm Q_{(l)} \Theta\left( \sum_{\bm Q'} | \tilde{s}^{\bm Q_{(l)}}_{\bm Q'} |\right).
\end{equation}
Here $\bm Q' \ge \max\{\bm 0,\bm Q_{(l)} - \bm q_i\}$ should be understood to ensure $\bm Q' -\bm Q_{(l)}+ \bm q_i \in \mathbb{N}^N$. Again multiplicities of inequivalent spurion classes are labeled by ${\bm Q}'$ and $\Theta$ is the Heaviside step function.
Then the same argument as in the non-supersymmetric case of \SEC{non-susy} can be applied, from which one can determine physical CP phases after diagonalizing the charge matrix and rewriting the scalar potential in the standard form of \EQ{standard-pot-multi}.

\subsection{Proof of the spurion mapping relation} \label{sec:proof}
Let us prove the relation~\eqref{Vspurion}.
The essential step is to unzip spurions from $\frac{\partial W}{\partial \phi_i}$ explicitly.
To express $\frac{\partial W}{\partial \phi_i}$  for a general superpotential $W$ and extract the spurions inside,
it is useful to first factorize the specific form of a superpotential $W$ out of the general set of monomial basis $\mathcal{F}_{\bm Q}$,
\begin{align}
  W  =\sum_{\bm{Q}  \in \mathbb{N}^N}  s^{\bm{Q}} \mathcal{F}_{\bm{Q}} (\Phi)= \hat{\bm{W}}_\Phi 
     \sum_{\bm{Q} \in \mathbb{N}^N}\mathcal{F}_{\bm{Q}} (\Phi)
\label{Woperator} \ ,
\end{align}
where $\hat{\bm{W}}_\Phi$ is the operator that maps a monomial basis to the corresponding term in the superpotential $W$, defined as $\hat{\bm{W}}_\Phi :\mathcal{F}_{\bm{Q}}(\Phi) \mapsto s^{\bm{Q}} \mathcal{F}_{\bm{Q}} (\Phi)$.
To construct  $\hat{\bm{W}}_\Phi$ explicitly, it is helpful to introduce the Dirac notation representing those monomial basis $\mathcal{F}_{\bm Q}$ as ket-states,
\begin{equation}
    |\bm Q\rangle_\Phi = \mathcal{F}_{\bm{Q}} (\Phi) \ .
\end{equation}
The dual state $\langle \bm Q|_\Phi$ that is canonically normalized can be constructed as 
\begin{equation}
\bra{\bm{Q}}_\Phi = \prod_{i=1}^{N}\left( \frac{1}{\bm Q_i!} \frac{\partial^{\bm Q_i}}{\partial \Phi_i^{\bm Q_i}} \right) \bigg|_{\Phi_i =0 } \ , 
\end{equation}
which satisfies $\langle{\bm Q | \bm Q'}\rangle_\Phi = \delta_{\bm{QQ}'}$.
Then the operator $\hat{\bm{W}}_\Phi$ simply reads
\begin{equation}
    \hat{\bm{W}}_\Phi = \sum_{\bm Q\in \mathbb{N}^N} s^{\bm Q} |\bm Q\rangle_{\Phi} \langle\bm Q|_\Phi \ .
    \label{opW}
\end{equation}
It is also straightforward to define the derivatives of states,
\begin{equation}
    \ket{\partial_i \bm Q}_\Phi  = \dfrac{\partial \mathcal{F}_{\bm Q }(\Phi)}{\partial \Phi_i} = \bm Q_i \ket{\bm Q - \bm q_i}_\Phi,
\end{equation}

\begin{equation}
    \bra{\partial_i \bm Q}_\Phi  =
    \begin{cases}
         \dfrac{1}{\bm Q_i} \bra{\bm Q - \bm q_i}_\Phi, \quad \bm Q_i\ne 0\\
         0, \qquad \qquad \qquad \,\, \bm Q_i =0
    \end{cases},
\end{equation}
where again $\bm q_i=(0,...,0,\underset{i}{1},0,...,0)$  is the unit charge vector corresponding to $U(1)_i$, and $\bm Q_i$ is the $i$-th component of $\bm Q$.
In analogy to \EQ{opW}, the derivative of $\bm{\hat W}$ reads
\begin{equation}
     \dfrac{\partial \bm{\hat  W}}{\partial\phi_i} = \sum_{\bm Q\in \mathbb N^N, \bm Q_i\ne 0} s^{\bm Q}  \ket{\bm Q - \bm q_i}_\phi \bra{\bm Q - \bm q_i}_\phi.
\end{equation}
Once the explicit form of $\hat{\bm{W}}_\Phi$ is obtained, one can calculate \EQ{mapW2V1} by using the derived operator $\frac{\partial \bm{\hat W}}{\partial\phi_i}$.
A monomial basis charged $\bm Q$ in the scalar potential is of the form $\mathcal{F}_{\bm{Q}'}(\phi) \mathcal{F}_{\bm{Q}'-\bm{Q}}^*(\phi) $.
Here $\bm Q'\ge\bm Q$ is required such that $\bm Q' - \bm Q \in \mathbb N^N$.
Therefore
\begin{eqnarray}
        V(\phi, \phi^*) &=&  \sum_{i=1}^N 
    \dfrac{\partial \bm{\hat  W}}{\partial\phi_i}  
    \left( \dfrac{\partial \bm{\hat  W}}{\partial\phi_i} \right)^*
    \sum_{\bm Q'\in \mathbb{N}^N, \bm Q'\ge\bm Q} \mathcal{F}_{\bm{Q}'}(\phi) \mathcal{F}_{\bm{Q}'-\bm{Q}}^*(\phi)  \nonumber\\
    &=&\sum_{i=1}^N   \sum_{\bm Q'\in \mathbb{N}^N, \bm Q'\ge\bm Q} 
    \left(  \dfrac{\partial \bm{\hat  W}}{\partial\phi_i} \ket{\bm Q'}_\phi \right)  
     \left( \dfrac{\partial \bm{\hat  W}}{\partial\phi_i} \ket{\bm Q'-\bm Q}_\phi \right)^*    \nonumber\\
      &=& \sum_{i=1}^N \sum_{\bm Q'\in \mathbb{N}^N, \bm Q' \ge\bm Q}   
      \left(s^{\bm{Q}'+\bm{q}_i} \mathcal{\tilde{F}}_{{\bm{Q}'} }(\phi) \right)
      \left(s^{\bm{Q}'-\bm{Q}+\bm{q}_i} \mathcal{\tilde{F}}_{\bm{Q}'-\bm{Q}}(\phi)\right)^*.\label{mapW2V3}
\end{eqnarray}
Eventually, \EQ{Vspurion} is proved by comparing \EQ{mapW2V2} and \EQ{mapW2V3}.

\subsection{R-charge analysis: radial direction stabilization}
R-symmetry plays a distinguished role in supersymmetric theories as the only continuous internal symmetry that acts nontrivially on the supercharges, thereby imposing strong constraints on the structure of the superpotential. In particular, the requirement that the superpotential carry R-charge two, $R(W)=2$, severely restricts the allowed operators and has important implications for the vacuum structure. A central result in this context is the Nelson–Seiberg theorem \cite{Nelson:1993nf}, which states that, for a generic superpotential, the presence of a continuous R-symmetry is necessary for spontaneous supersymmetry breaking, while its spontaneous breaking is required for phenomenologically viable SUSY-breaking vacua. It is therefore natural to discuss SCPV within an R-symmetric supersymmetric framework.

Since we are interested in SCPV vacua stabilized in the exact SUSY limit, we must require that the R-symmetry remain unbroken in the CP-violating vacua (although it may be spontaneously broken in other sectors). This requirement implies that all superfields carrying nonzero R-charge must have vanishing vacuum expectation values. As we show below, this condition leads to nontrivial constraints on the form of the superpotential.

Consider a theory with $N$ chiral superfields. We denote by $\Phi_\alpha^{(r)}$ the 
$\alpha$-th superfield carrying R-charge  $r$, where $1 \leq \alpha \leq N_r$ and $\sum_r N_r =N$. Since the superpotential carries R-charge two, we formally denote it by $W^{(2)}$. In a SUSY-preserving vacuum, all F-term conditions must be satisfied, 
\begin{equation}
    \frac{\partial W^{(2)}}{\partial \Phi_\alpha^{(r)}} = 0,
\end{equation}
for all values of $r$ and $\alpha$. Each such equation carries R-charge $2-r$.

According to the Nelson–Seiberg theorem, in a theory with a continuous R-symmetry, spontaneous breaking of the R-symmetry implies spontaneous breaking of supersymmetry. Therefore, in SUSY-preserving vacua, all R-charged superfields must have vanishing VEVs. As a consequence, any F-term equation with nonzero R-charge must trivially vanish identically at the vacuum, since its left-hand side necessarily contains at least one R-charged superfield. While the existence of such solutions may require additional tuning of parameters, we are here concerned only with necessary conditions.

The only potentially nontrivial F-term equations that remain are those that are R-neutral, which arise from derivatives with respect to R-charge–two superfields,
\begin{equation}
    \frac{\partial W^{(2)}}{\partial \Phi_\alpha^{(2)}} = 0. 
\end{equation}
Even among these equations, the left-hand side may still involve R-charged superfields—for instance, through bilinears of fields carrying opposite R-charges—which again vanish upon imposing zero VEVs for all R-charged fields. Such equations can be systematically identified and discarded. After this amputation procedure, suppose that we are left with $n_2$ independent F-term equations whose left-hand sides involve only $n_0$ R-neutral superfields (up to constant terms).

A necessary condition for stabilizing the R-neutral fields in a supersymmetric vacua is then
\begin{equation}
    n_2 \geq n_0,
\end{equation}
namely that the number of independent equations is at least as large as the number of variables. More precisely, $n_2$ should be interpreted as the number of R-charge–two superfields that possess at least one interaction term involving only R-neutral fields, while $n_0$ counts the R-neutral superfields that couple exclusively to such R-charge–two fields, together with other R-neutral fields.

It is important to emphasize that this stabilization condition derived from R-charge analysis is insensitive to CP-violating phases. In this sense, it constrains only the stabilization of radial directions in field space. To diagnose the existence of physical CP-violating phases, this analysis must therefore be supplemented by the spurion approach discussed in subsequent sections. Moreover, the above condition does not guarantee the existence of vacua with nonzero VEVs for the R-neutral fields; it merely provides a necessary condition for vacuum stabilization, rather than for the existence of isolated vacua free of flat directions.

Finally, we stress that this argument is completely general. No assumption has been made regarding the renormalizability of the theory, and the above conclusions apply equally to superpotentials containing arbitrary non-renormalizable operators.

\subsection{An example
\label{sec:example}}
We now illustrate our general framework by revisiting an example originally discussed in Ref.~\cite{Dine:2015jga}, in which CP is spontaneously broken in isolated vacua due to supersymmetric dynamics.
Applying the procedure developed in \SEC{sec:susycase}, one can explicitly verify that our formalism correctly captures the necessary conditions for SCPV in this model.

The setup of Ref.~\cite{Dine:2015jga} consists of four chiral superfields, $X,Y, \eta_1, \eta_2$, with the superpotential
\begin{equation}
    W = X \mu^2 +X(a \eta_1^2+b\eta_1 \eta_2+c\eta_2^2) + Y(a'\eta_1^2+b'\eta_1 \eta_2 +c'\eta_2^2) \ ,
    \label{example}
\end{equation}
where $\mu,a,a',b,b',c,$ and $c'$ are real parameters. The superfields $X$ and $Y$ carry R-charge two and are even under $Z_2$, while $\eta_{1,2}$ are R-neutral and odd under $Z_2$. This superpotential generically admits supersymmetric vacua in which $\eta_1$ and $\eta_2$ acquire complex vacuum expectation values, thereby spontaneously breaking CP. 

\renewcommand{\arraystretch}{1.5} 
\begin{table}[t!] 
\begin{tabular}{|c|c|}
  \hline
    Inequivalent spurion class $ \tilde{s}^{\bm Q}$& Charge  $\bm{Q}$\\
    \hline
     $a^* \mu^2$ &  $\pm(2, 0, 0, 0)$ \\
    \hline
     $b^* \mu^2$ &  $\pm(1, 1, 0, 0)$ \\
    \hline
     $c^* \mu^2$ &  $\pm(0, 2, 0, 0)$ \\
     \hline
     $a^*a' , 2 b^*b' , c^*c'$& $\pm(0, 0, 1, -1)$ \\
     \hline
     $2 a^*b ,2 b^*c ,2  (a')^*b' ,2 (b')^*c'$& $\pm (1,-1,0,0)$ \\
     \hline
     $a^*b',b^*c'$& $\pm (1, -1,1,-1)$ \\
     \hline
     $(a')^*b,(b')^*c$& $\pm (1, -1,-1,1)$ \\
     \hline
     $a^*c,(a')^*c'$& $\pm(2,-2,0,0)$ \\
     \hline
\end{tabular}
\centering
\vspace{0.3cm}
\caption{The summary of spurions in the scalar potential and their corresponding charges
for the superpotential given in \EQ{example}.}
\label{Tab_2}
\end{table}

We begin by analyzing the stabilization of CP-violating phases using the spurion formalism. For notational convenience, we relabel the fields $\eta_1, \eta_2, X$, and $Y$ as $\Phi_1, \Phi_2, \Phi_3$, and $\Phi_4$, respectively, so that the charge vector is fixed.
The parameters appearing in \EQ{example} are treated as superpotential spurions $s^{\bm{Q}}$. For example, the parameter $a$ corresponds to $s^{(2,0,1,0)}$ and $\mu^2$ corresponds to $s^{(0,0,1,0)}$.
Then, we apply the relation in \EQ{Vspurion} to map the superpotential spurions $s^{\bm{Q}}$ to the scalar potential spurions $\tilde{s}^{\bm{Q}}_{\bm Q'}$.
With the help of our calculation program, which is shown in Appendix \ref{sec:algorithm}, we find that there are 8 inequivalent spurions of the scalar potential, as listed in Table~\ref{Tab_2}. The charge matrix $\mathcal{Q}$ is then identified as 
\begin{equation}\label{exampleQ}
   \mathcal{Q}= \begin{pmatrix}
-2 & 0 & 0 & 0 \\[-2ex]
-1 & -1 & 0 & 0 \\[-2ex]
0 & 0 & -1 & 1 \\[-2ex]
0 & -2 & 0 & 0 \\[-2ex]
-1 & 1 & 0 & 0 \\[-2ex]
-1 & 1 & -1 & 1 \\[-2ex]
-1 & 1 & 1 & -1 \\[-2ex]
-2 & 2 & 0 & 0
\end{pmatrix},
\end{equation}
up to an overall minus sign on each row. Note that the last column is identical with the third column up to a minus sign, so that we can  truncate it. Then taking the right pseudo-inverse of $\mathcal{Q}$, which is the inverse matrix of the top-left $3 \times 3$ matrix in \EQ{exampleQ}, we obtain
\begin{equation}
\mathcal{Q}' =
\begin{pmatrix}
\mathcal{Q}_{3\times 3}^{\text{break}} & \bm{0}_{3\times 1} \\
 \mathcal Q_{5\times 3}^{\text{support}} & \bm{0}_{5\times 1}        
\end{pmatrix}
=\begin{pmatrix}
1 & 0 & 0 & 0 \\[-2ex]
0 & 1 & 0 & 0 \\[-2ex]
0 & 0 & 1 & 0 \\[-2ex]
-1 & 2 & 0 & 0 \\[-2ex]
1 & -1 & 0 & 0 \\[-2ex]
1 & -1 & 1 & 0 \\[-2ex]
1 & -1 & -1 & 0 \\[-2ex]
2 & -2 & 0 & 0
\end{pmatrix},
\end{equation}
as in the form of \EQ{Q'}.
Based on \EQ{nCPphase}, we find $d=3$. Thus one can conclude that the superpotential of \EQ{example} satisfies the necessary condition of SCPV in phase directions.

We next examine the R-charge constraints. In the present model, both $X$ and $Y$ have interaction terms involving only the R-neutral fields $\eta_1$ and $\eta_2$, and no additional R-charged superfields are present. Consequently, we have $n_2 = n_0=2$, and the necessary condition for stabilizing supersymmetric vacua is satisfied.

It is instructive to ask whether this construction is minimal and to illustrate how the two necessary conditions constrain model building. Let us first consider removing one of the R-neutral fields, say $\eta_2$. Although the original supersymmetric Nelson–Barr realization requires two $Z_2$-odd superfields, we temporarily set this aside and focus solely on SCPV. Setting $b=c=b'=c' =0$, the superpotential reduces to
\begin{equation} \label{cancel2}
    W_{\cancel{\eta_2}} = X\mu^2+X(a\eta_1^2) + Y\lambda^2+Y(a'\eta_1^2),
\end{equation}
where we have included the term $Y\lambda^2$ with real parameter $\lambda$ without loss of generality. The resulting scalar spurions are listed in Table~\ref{Tab_3}. In this case, the number of inequivalent spurions equals the rank of the charge matrix, $N_s = r=2$, implying a trivial supporting matrix and hence $d=0$. The necessary condition for SCPV in the phase direction is therefore not satisfied.

Alternatively, one may keep both $\eta_1$ and $\eta_2$ but remove one of the R-charged fields, say $Y$. Setting $a'=b'=c'=0$, the superpotential becomes
\begin{equation} \label{cancelY}
    W_{\cancel{Y}} = X\mu^2 + X(a\eta_1^2+b\eta_1 \eta_2 + c\eta_2^2).
\end{equation}
The spurion analysis alone does not exclude this case, as shown in Table~\ref{Tab_4}. However, the R-charge analysis immediately rules it out: with a single R-charge–two superfield coupled to two R-neutral fields, one has $n_2 < n_0$, so the necessary condition for stabilizing supersymmetric vacua fails. Consequently, either supersymmetry is spontaneously broken or the vacuum contains a flat direction.

Combining both the spurion and R-charge analyses, we conclude that the superpotential \EQ{example} constitutes the minimal realization of spontaneous CP violation in a supersymmetric vacuum consistent with the imposed symmetries. Together, these two necessary conditions provide a powerful and systematic guide for model building in SUSY-conserving realization of SCPV.

\renewcommand{\arraystretch}{1.5} 
\begin{table}[t!] 
\begin{tabular}{|c|c|}
  \hline
    Inequivalent spurion class $ \tilde{s}^{\bm Q}$& Charge  $\bm{Q}$\\
    \hline
     $a^* \mu^2, (a')^*\lambda^2 $ &  $\pm(2, 0, 0, 0)$ \\
    \hline
     $a^*a' $& $\pm(0, 0, 1, -1)$ \\
     \hline
\end{tabular}
\centering
\vspace{0.3cm}
\caption{The summary of spurions in the scalar potential and their corresponding charges
for the superpotential given in \EQ{cancel2}.}
\label{Tab_3}
\end{table}

\renewcommand{\arraystretch}{1.5} 
\begin{table}[t!] 
\begin{tabular}{|c|c|}
  \hline
    Inequivalent spurion class $ \tilde{s}^{\bm Q}$& Charge  $\bm{Q}$\\
    \hline
     $a^* \mu^2$ &  $\pm(2, 0, 0, 0)$ \\
    \hline
     $b^* \mu^2$ &  $\pm(1, 1, 0, 0)$ \\
    \hline
     $c^* \mu^2$ &  $\pm(0, 2, 0, 0)$ \\
     \hline
     $2 a^*b ,2 b^*c$& $\pm (1,-1,0,0)$ \\
     \hline

     $a^*c$& $\pm(2,-2,0,0)$ \\
     \hline
\end{tabular}
\centering
\vspace{0.3cm}
\caption{The summary of spurions in the scalar potential and their corresponding charges
for the superpotential given in \EQ{cancelY}.}
\label{Tab_4}
\end{table}

\section{SCPV on pseudo-flat directions}
\label{susy-breaking case}

We here explore a scenario where CP symmetry is spontaneously broken on pseudo-flat directions lifted
by soft SUSY breaking and non-perturbative effects, predicting particles at the soft SUSY breaking scale. 
In \SEC{sec:setup}, we begin with the introduction of our setup, and
\SEC{sec:stabilization} discusses the stabilization of a vacuum with a physical CP-violating phase.
In \SEC{sec:CKM}, we embed our model into the Nelson-Barr framework to address the strong CP problem.

\subsection{Setup
\label{sec:setup}}

Supersymmetric theories generally possess flat directions (for e.g. the minimal supersymmetric SM (MSSM) and next-to-minimal supersymmetric SM (NMSSM), see Refs.~\cite{Luty:1995sd,Dine:1995kz,Gherghetta:1995dv,Fayet:1974pd}, and for the existence of non-Goldstone classically-flat directions in SUSY theories, see Refs.~\cite{Fayet:1975ki}).
We pursue a possibility that CP symmetry is spontaneously broken in such field spaces. 
To find a concrete example, let us consider a model with the following superpotential:
\begin{equation}
    W = \lambda X (\Phi_1 \Phi_2 - v^2) \ ,
    \label{W_SUSY}
\end{equation}
where $\lambda$ and $v$ are real parameters with mass dimension zero and one, respectively, and $X$, $\Phi_1$ and $\Phi_2$ are chiral superfields.
The superpotential exhibits a (spurious) $U(1)$ symmetry under which $\Phi_1$ and $\Phi_2$ have charges with the same absolute value but opposite signs. The $U(1)$ symmetry is explicitly broken by SUSY breaking effects as well as non-perturbative effects of some strong gauge dynamics, as will be introduced shortly.\footnote{If the $U(1)$ symmetry is only broken by non-perturbative effects of QCD,  the superpotential (\ref{W_SUSY}) has been well-studied in the context of SUSY axion models (see e.g. Refs.~\cite{Kim:1983ia,Rajagopal:1990yx,Kasuya:1996ns,Covi:2001nw,Kawasaki:2007mk,Kawasaki:2010gv,Kawasaki:2011ym,Bae:2011jb,Bae:2011iw,Nakayama:2012zc,Moroi:2012vu,Kawasaki:2013ae,Ema:2017krp,Co:2017mop,Ema:2018abj,Ema:2021xhq,Nakai:2021nyf,Co:2023mhe}).}
The corresponding scalar potential is then given by
\begin{equation}
    V_{F} =  \lambda^2|\phi_1 \phi_2 - v^2|^2 + \lambda^2 |X|^2 (|\phi_1|^2 + |\phi_2|^2) \ ,
    \label{VF}
\end{equation}
where $\phi_{1,2}$ denote the scalar components of $\Phi_{1,2}$, respectively.
Thus, the VEVs for $X, \phi_1,\phi_2$ are
\begin{equation}
    \expval{X} = 0, \ \ \ \expval{\phi_1} = v_1 e^{i\theta}, \ \ \ \expval{\phi_2} = v_2 e^{-i\theta},
    \label{vev}
\end{equation}
where $v_1v_2 \equiv v^2$, and $\theta \in[0,2\pi)$ is an arbitrary parameter.
The fact that $\theta$ and $v_1$ (or $v_2$) are not uniquely determined implies the existence of flat directions,
which correspond to massless fields. 
We assume that those flat directions are not lifted by $U(1)$ breaking terms in the tree-level superpotential.
Instead, explicit $U(1)$ breaking terms are provided by soft SUSY breaking, and their scale is much lower than that of SCPV $m_{\rm CP} ~ (\equiv v)$.
This hierarchy is motivated by the suppression of radiative corrections to the strong CP phase  \cite{Dine:1993qm,Dine:2015jga,Hiller:2001qg}.

When the scalar fields $\phi_{1,2}$ develop complex VEVs $(\theta\neq0,\pi)$, CP symmetry is spontaneously broken.
In the vicinity of the VEVs given by \EQ{vev}, $\phi_1$ and $\phi_2$ are generally expanded in the forms of
\begin{align}
    &\phi_1(x) 
    = \lmk v_1+\frac{\sigma_1(x)}{\sqrt{2}}\rmk \exp\left[i\lmk \theta+\frac{\pi_1(x)}{\sqrt{2}v_1}\rmk\right], \label{Phi1} \\[1.5ex]
    &\phi_2(x)
    = \lmk v_2+\frac{\sigma_2(x)}{\sqrt{2}}\rmk \exp\left[i\lmk-\theta+\frac{\pi_2(x)}{\sqrt{2}v_2}\rmk\right]. \label{Phi2}
\end{align}
Diagonalizing the mass matrices for $\sigma_1,\sigma_2$ and $\pi_1,\pi_2$, we can identify the massless modes as follows:
\beq
s(x) &=& \frac{1}{f_a} (v_1\sigma_1(x)-v_2\sigma_2(x)) \ ,\\[1ex]
a(x) &=& \frac{1}{f_a}(v_1\pi_1(x)-v_2\pi_2(x)) \ ,
\label{axion}
\eeq
with  $f_a\equiv\sqrt{v_1^2+v_2^2}$ defined as the normalization factor.
In the context of axion physics, these mass eigenstates can be identified as {\it saxion} and {\it axion} with the decay constant $\sim f_a$.
In addition, there are also fermion partners, {\it axino}, which remain massless in the SUSY limit.

Physical CP-violating phases should be determined by terms which break the spurious $U(1)$ symmetry. 
One needs at least two such terms with different periodicity, e.g., $V = \mathcal{C}\cos\theta+\mathcal{D}\cos(2\theta)$ (see \SEC{non-susy}).
We consider two sources of the $U(1)$ breaking originated from soft SUSY breaking and non-perturbative effects of a gauge theory.

\subsubsection{Soft SUSY breaking
\label{sec:SUSYbreaking}}

First let us introduce soft SUSY breaking terms, given by
\begin{align} \label{soft}
    V_{\rm soft} &= 
    \lmk \frac{1}{2}b_1 \phi_1^2 + \frac{1}{2}b_2 \phi_2^2  + {\rm h.c.}\rmk
    + m_1^2\phi_1^*\phi_1+m_2^2\phi_2^*\phi_2 \ ,
\end{align}
where $b_{1,2}$, $m^2_{1,2}$ denote constant parameters with mass dimension 2 
and are assumed to be much smaller than $m_{\rm CP}^2$.
One can see that the $b$-terms induce potentials proportional to $\cos(2\theta)$ by using Eqs.~(\ref{Phi1}), (\ref{Phi2}), while the latter give constant terms for $\theta$.
In addition to \EQ{soft}, we can also have other terms, like mixing terms of $\phi_1,\  \phi_2$ and $A$-terms. 
The $A$-terms, e.g. $\phi_1^2\phi_2$, can induce potential terms $\propto \cos\theta$ which might stabilize a CP-violating vacuum together with Eq.~(\ref{soft}).
However, in this case, the other $A$-terms such as $\phi_1^3+\phi_2^3$ are also possible and induce the dominant contribution in the limit of $v_1\rightarrow0$ or $\infty$, because these terms give the dependence like $v_1^3+v^6/v_1^3$, which is stronger than the other terms.
Thus, the total potential would be unbounded from below.
When discussing the Nelson-Barr setup in \SEC{sec:CKM}, we will require a $Z_2$ symmetry which forbids all the $A$-terms and even other trilinear terms such as $\phi_1\phi_2^{*2}$.
Although other bilinear terms such as $\phi_1\phi_2, \ \phi_1\phi_2^*$ cannot be forbidden, they provide the same shape in the phase direction, i.e. $\cos(2\theta)$ or constant term, and do not alter the final results qualitatively. 
Therefore, we simply assume the form of Eq.~(\ref{soft}) for soft SUSY breaking.

Note that the last two terms in Eq.~(\ref{soft}) are essential for obtaining a stable vacuum on the flat direction.
By using the parameterization of Eqs.~\eqref{Phi1}, \eqref{Phi2} on the flat direction, the potential (\ref{soft}) is rewritten as
\beq
V_{\rm soft} = \lmk b_1v_1^2 + b_2\frac{v^4}{v_1^2}\rmk \cos(2\theta) +m_1^2v_1^2 +m_2^2\frac{v^4}{v_1^2} \ .
\label{soft_vac}
\eeq
In the limit of $v_1\rightarrow\infty$ or $0$, the potential is bounded from below under the condition that
\beq
|b_1| \leq m_1^2 \ ,~~|b_2| \leq m_2^2 \ . \label{bounded_cond}
\eeq
When this condition is satisfied, the potential is minimized at $\theta=\pi/2$, but this solution can be shifted by $\theta\rightarrow\theta-\pi/2$ or $b_1,\; b_2\rightarrow -b_1,\; -b_2$.
Therefore, other terms which are not proportional to $\cos(2\theta)$ are required to obtain a CP-violating vacuum.

\subsubsection{Non-perturbative effects}

We now consider a supersymmetric $SU(N)$ gauge theory with a vector-like pair of (anti-)quark supermultiplets
$Q, \, \Bar{Q}$ which transform as fundamental and anti-fundamental representations under $SU(N)$.
They couple to $\Phi_{1,2}$ through the following superpotential,\footnote{When $\kappa_2=0$,
the $U(1)$ symmetry is explicitly broken only through anomaly.
However, together with an $R$ symmetry, we can still define an anomaly-free $U(1)$ symmetry.
Thus, a nonzero potential for the phase direction is induced by some explicit $R$ symmetry breaking, such as the gaugino mass.
This scenario is potentially viable but not further pursued in the present paper.}
\begin{equation}
    W_Q = (\kappa_1 \Phi_1 + \kappa_2 \Phi_2) \bar{Q} Q \ ,
    \label{WQ}
\end{equation}
where $\kappa_1,\kappa_2$ are real coupling constants. 
Since $\phi_1$ and/or $\phi_2$ develop the nonzero VEVs, $Q,\bar{Q}$ are decoupled at the scale of $m_{\rm eff} \equiv \kappa_1\phi_1+\kappa_2 \phi_2$.
After the decoupling, the theory turns into the pure SUSY Yang-Mills, and the gaugino condensation provides the effective superpotential as
\beq
W_{\rm eff} = N\Lambda_{\rm eff}^3 \ ,
\label{Weff}
\eeq
with 
\beq
\Lambda_{\rm eff}^3 = \lmk\frac{m_{\rm eff}}{\Lambda}\rmk^{1/N}\Lambda^{3} \ .
\label{dynamical}
\eeq
Here $\Lambda$, $\Lambda_{\rm eff}$ respectively denote the holomorphic dynamical scales before and after the decoupling, and the relation (\ref{dynamical}) is given by the matching condition of the $SU(N)$ gauge coupling constant at the decoupling scale.

The spurious $U(1)$ symmetry is explicitly broken by $\Phi_2\bar{Q}Q$ (or $\Phi_1\bar{Q}Q$) in \EQ{WQ}, which induces a  potential in the phase direction.
The dynamically generated superpotential $W_{\rm eff}$ contributes to the scalar potential, 
\begin{align}
    V_{\rm dyn} = \left| \frac{\partial W_{\text{eff}}}{\partial \phi_1} \right|^2 + \left| \frac{\partial W_{\text{eff}}}{\partial \phi_2} \right|^2 
=\frac{(\kappa_1^2+\kappa_2^2)\,\Lambda^{6-\frac{2}{N}}}{|\kappa_1 \phi_1+\kappa_2\phi_2|^{2-\frac{2}{N}}} \ .
\label{Vdyn}
\end{align}
As $\phi_1$ and $\phi_2$ develop VEVs, the scalar potential can reduce to 
\begin{equation}
    V_{\rm dyn} = \frac{(\kappa_1^2+\kappa_2^2)\Lambda^{6-\frac{2}{N}}}{[\kappa_1^2v_1^2+\kappa_2^2 v^4/v_1^2+2\kappa_1\kappa_2v^2\cos(2\theta)]^{1-\frac{1}{N}}} \ .
    \label{Vdyn_vac}
\end{equation} 
This potential is obviously positive-definite and then bounded from below. 
We will shortly see that the combination of this non-perturbatively generated potential and the soft SUSY breaking terms stabilizes a vacuum with a nonzero physical complex phase.

\subsection{Vacuum stabilization
\label{sec:stabilization}}

Combining all the three scalar potential contributions (\ref{VF}), (\ref{soft}) and (\ref{Vdyn}), we can write down the total scalar potential on the flat direction,
\begin{align}
    V_{\rm tot} 
    &= V_{F} + V_{\rm soft} + V_{\rm dyn} \notag \\[1ex]
    &= b_1\lmk v_1^2 + \frac{b_2}{b_1} \frac{v^4}{v_1^2}\rmk\, \cos\lmk 2\theta\rmk +m_1^2 \lmk v_1^2 + \frac{m_2^2}{m_1^2} \frac{v^4}{v_1^2}\rmk \notag\\
    & \ \ \ +\frac{(\kappa_1^2+\kappa_2^2)\Lambda^{6-\frac{2}{N}}}{[\kappa_1^2v_1^2+\kappa_2^2 v^4/v_1^2+2\kappa_1\kappa_2v^2\cos(2\theta)]^{1-\frac{1}{N}}} \ .
    \label{Vtot}
\end{align}
Here $V_F=0$ as we assume the hierarchy of $m_{\rm CP} \gg m_{\text{soft}}$, or equivalently, $v^2 \gg |m_{1,2}^2|$.
There are two $\cos(2\theta)$ terms, and thus, the periodicity is $\pi$, but one is in the numerator and the other in the denominator, leading to a non-trivial CP violation.
Since both $b_i$ and $m_i^2$ originate from soft SUSY breaking, no significant hierarchy is expected between them.
It is therefore natural to consider $\alpha_b \equiv b_2 / b_1 = \mathcal{O}(1)$, $\alpha_m \equiv m_2^2 / m_1^2 = \mathcal{O}(1)$, and $\beta \equiv b_1 / m_1^2 = \mathcal{O}(1)$. For numerical convenience in exploring the viable parameter space, we use these dimensionless parameters to write down the following dimensionless potential:
\begin{align}
\tilde{V}_{\rm tot} \equiv \frac{V_{\rm tot}}{m_1^2v^2} &=\beta\lmk\tilde{v}_1^2 +\frac{\alpha_b}{\tilde{v}_1^2}\rmk\, \cos(2\theta) + \lmk\tilde{v}_1^2+\frac{\alpha_m}{\tilde{v}_1^2}\rmk \notag \\
& \ \ \ +\frac{(\kappa_1^2+\kappa_2^2)\, \gamma}{[\kappa_1^2\tilde{v_1}^2+\kappa_2^2/\tilde{v_1}^2+2\kappa_1\kappa_2\cos(2\theta)]^{1-\frac{1}{N}}} \ , \label{dimlessV}
\end{align}
in which we define $\tilde{v}_1^2 \equiv v_1^2/v^2$ and $\gamma \equiv \Lambda^{6-\frac{2}{N}}/(m_1^2v^{4-\frac{2}{N}})$.

For the potential to be bounded from below, we require that $|\beta| = |b_1|/m_1^2 \leq 1$, as indicated in Eq.~(\ref{bounded_cond}). 
In the following analysis, we take $\kappa_1 = \kappa_2 = 1$ and $\beta > 0$ as a simple benchmark set,
since there is only a tiny parameter space to correctly stabilize $\theta$ for $\beta<0$ and $\kappa_1\kappa_2>0$.
To see the reason for this, consider the stationary condition along the phase direction, which requires
\begin{align}
\frac{\partial \tilde{V}_\text{tot}}{\partial (\cos 2\theta)}
&= \beta \lmk \tilde{v}_1^2 + \frac{\alpha_b}{\tilde{v}_1^2} \rmk
- 2 \kappa_1 \kappa_2 \left(1 - \frac{1}{N}\right)
\frac{(\kappa_1^2 + \kappa_2^2)\gamma}{
[\kappa_1^2 \tilde{v}_1^2 + \kappa_2^2 / \tilde{v}_1^2 + 2\kappa_1\kappa_2\cos(2\theta)]^{2 - \frac{1}{N}}} \notag \\
&= 0  .
\end{align}
The sign of the second term is determined by $-\text{sgn}(\kappa_1 \kappa_2)$, while in the first term, $\lmk \tilde{v}_1^2 + \frac{\alpha_b}{\tilde{v}_1^2} \rmk$ is typically positive without fine-tuning.
Therefore, for $\beta > 0$ and $\kappa_1 \kappa_2 < 0$ or equivalently for $\beta < 0$ and $\kappa_1 \kappa_2 > 0$, we have $\frac{\partial \tilde{V}_\text{tot}}{\partial (\cos 2\theta)} \neq 0$, implying that the parameter space admitting stationary points is quite restricted.

\begin{figure}[!t]
  \centering
  \begin{subfigure}{0.32\textwidth}
    \includegraphics[width=\textwidth]{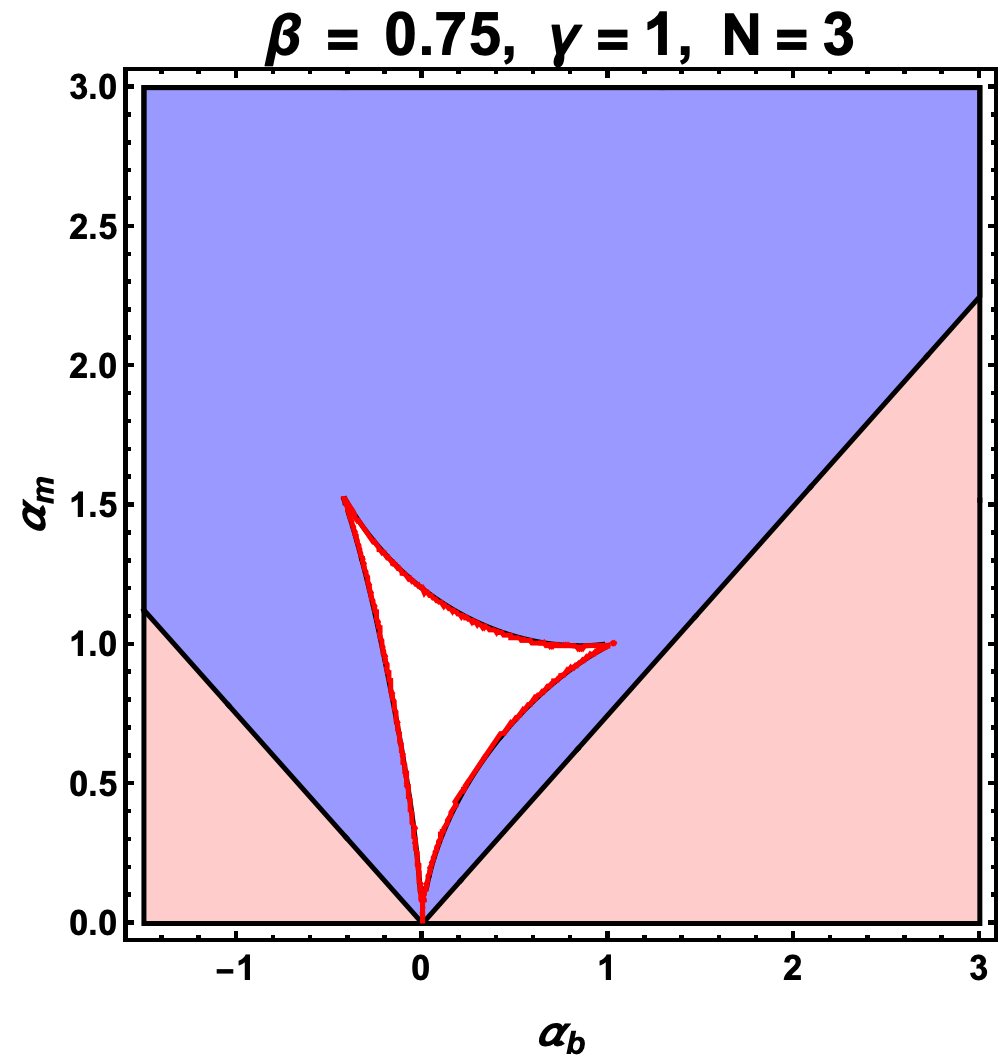}\caption{} \label{5-a}
  \end{subfigure}
  \begin{subfigure}{0.32\textwidth}
    \includegraphics[width=\textwidth]{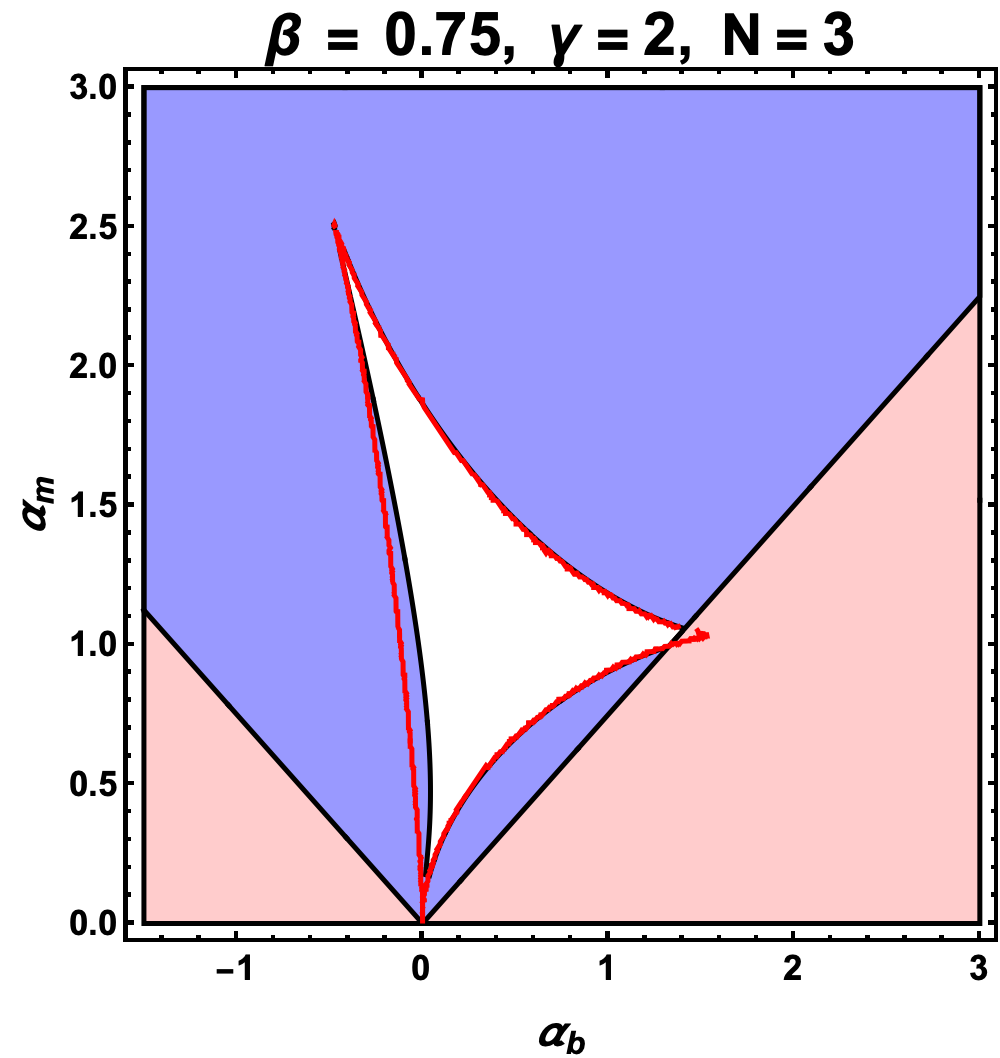}\caption{} \label{5-b}
  \end{subfigure}
  \begin{subfigure}{0.32\textwidth}
    \includegraphics[width=\textwidth]{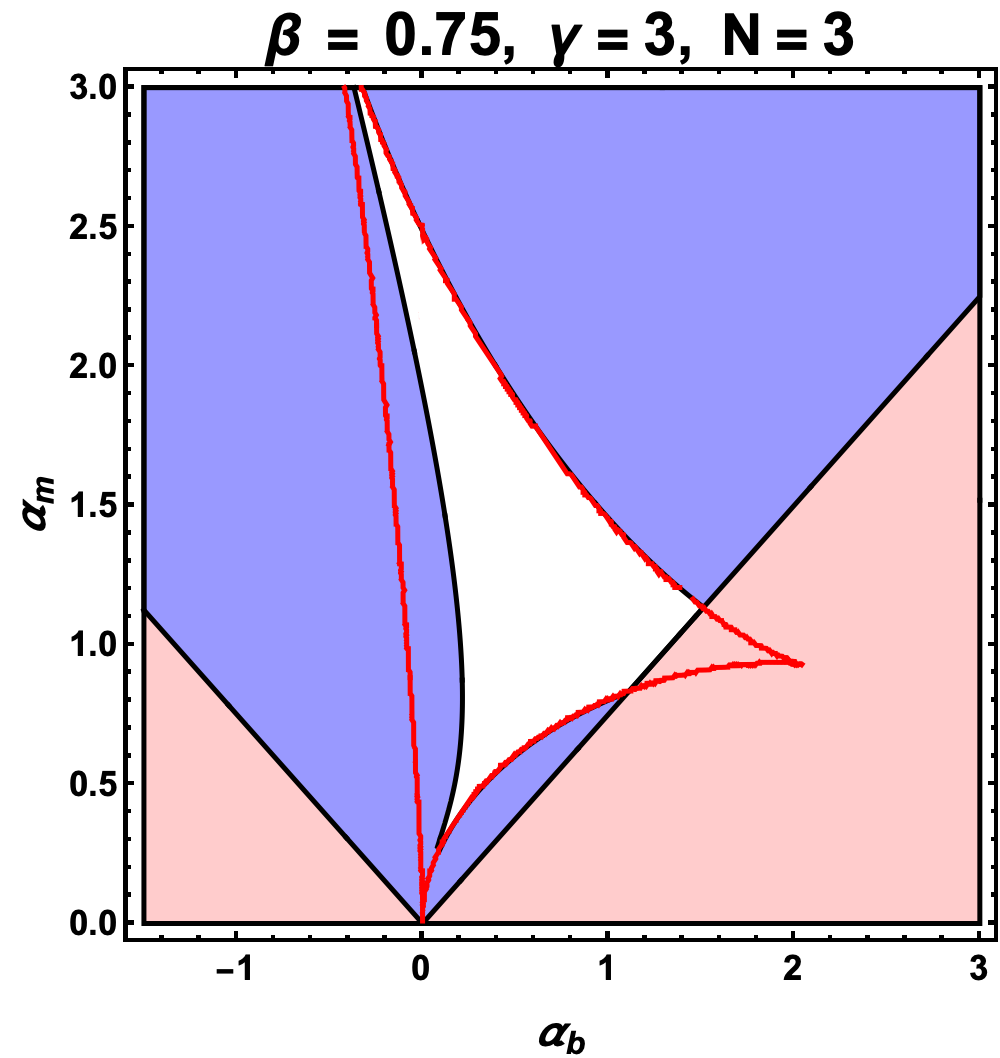}\caption{} \label{5-c}
  \end{subfigure}

  \vspace{0.5cm}

  \begin{subfigure}{0.32\textwidth}
    \includegraphics[width=\textwidth]{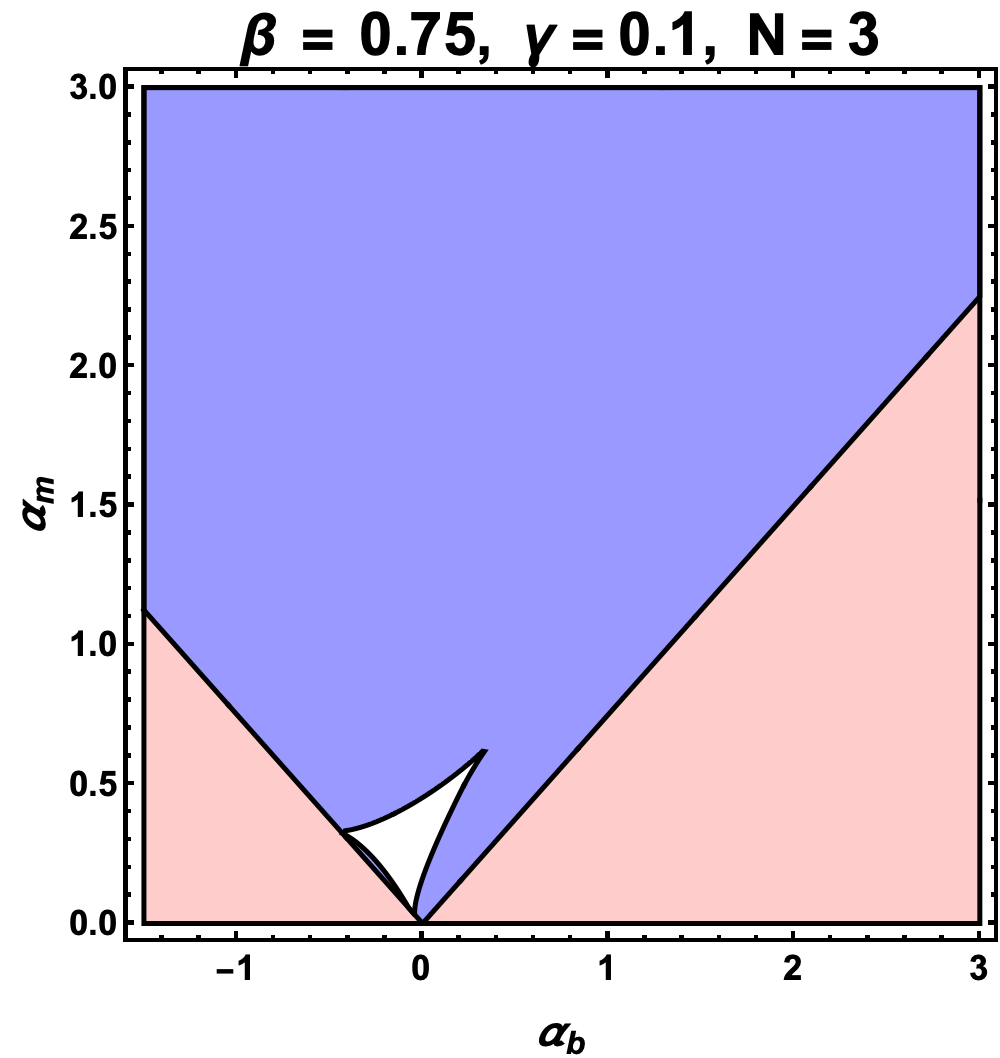}\caption{} \label{5-d}
  \end{subfigure}
  \begin{subfigure}{0.32\textwidth}
    \includegraphics[width=\textwidth]{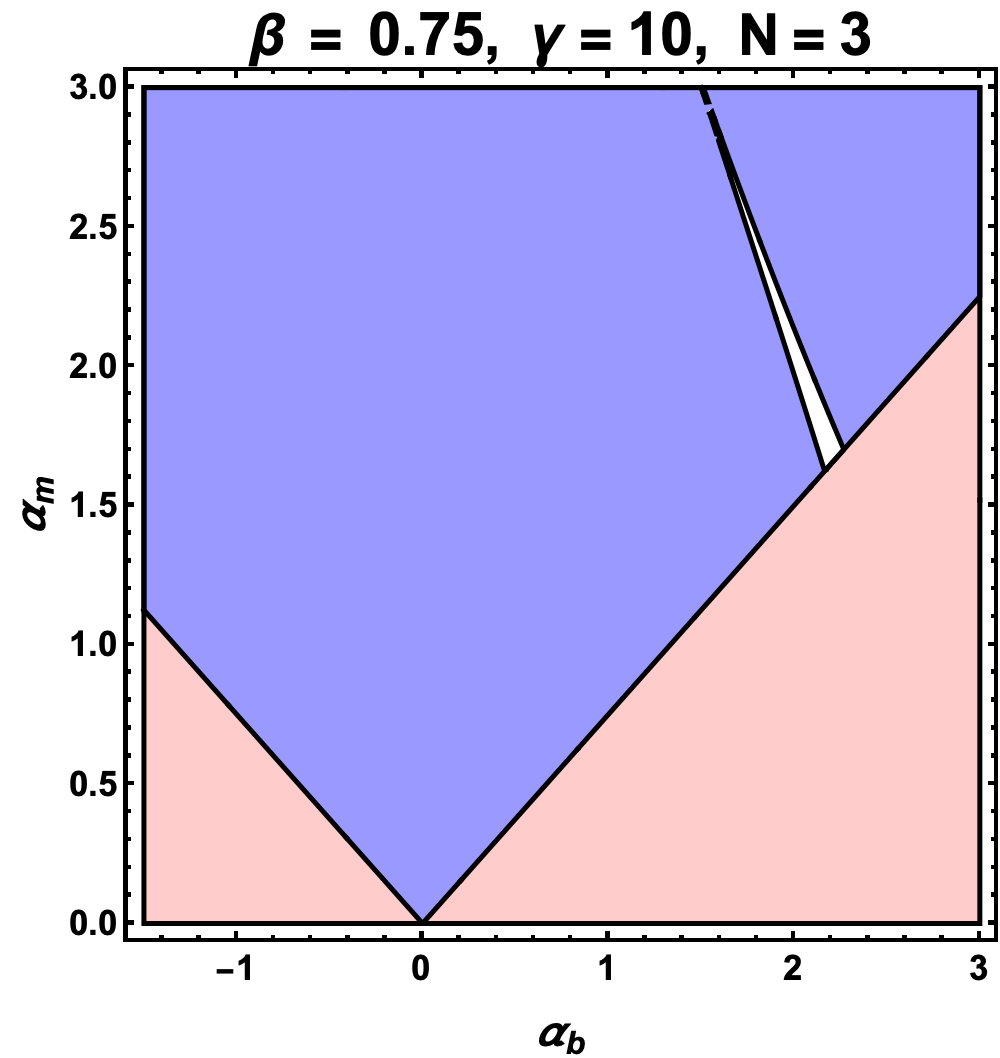}\caption{} \label{5-e}
  \end{subfigure}

  \caption{The parameter space in the plane of $\alpha_b$-$\alpha_m$ for $\gamma = 1, 2, 3,0.1$ and $10$ with $\beta = 0.75$, $N = 3$ and $\kappa_1 = \kappa_2 = 1$ fixed. 
  In the pink shaded regions, the potential is unbounded from below, given by \EQ{bounded_cond}, while in the blue shaded regions, the potential is bounded from below but SCPV is not obtained. 
  The white regions represent the viable parameter region that realizes correctly stabilized CP-violating vacua within a bounded-from-below potential. 
  The inside of the red curves in panels (a)–(c) indicates that the maximal determinant of the Hessian matrix at the stationary points is positive.}
  \label{fig:β=0.75,Ν=3}
\end{figure}

\begin{figure}[!t]
  \centering
  \begin{subfigure}[b]{0.32\textwidth}
    \centering
    \includegraphics[width=\textwidth]{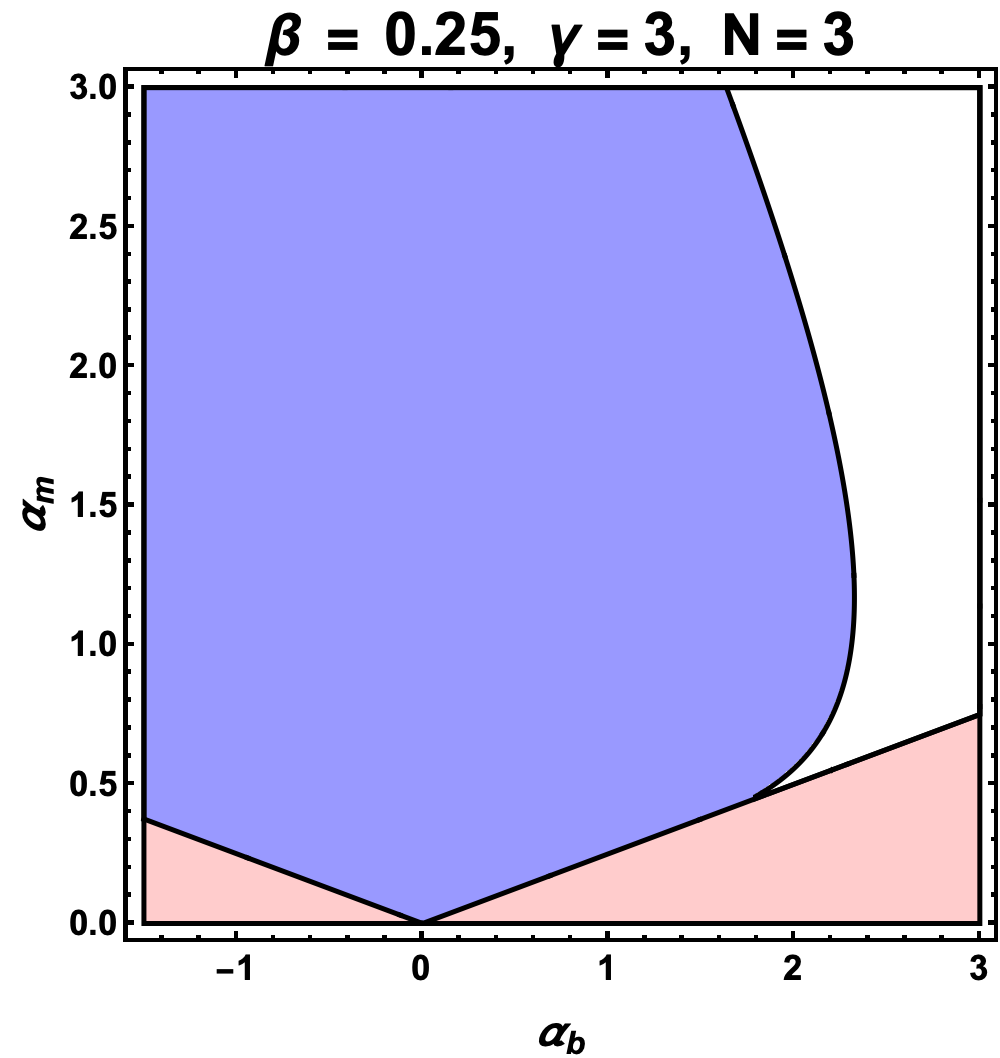}
    \caption{} \label{6-a}
  \end{subfigure}\hfill
  \begin{subfigure}[b]{0.32\textwidth}
    \centering
    \includegraphics[width=\textwidth]{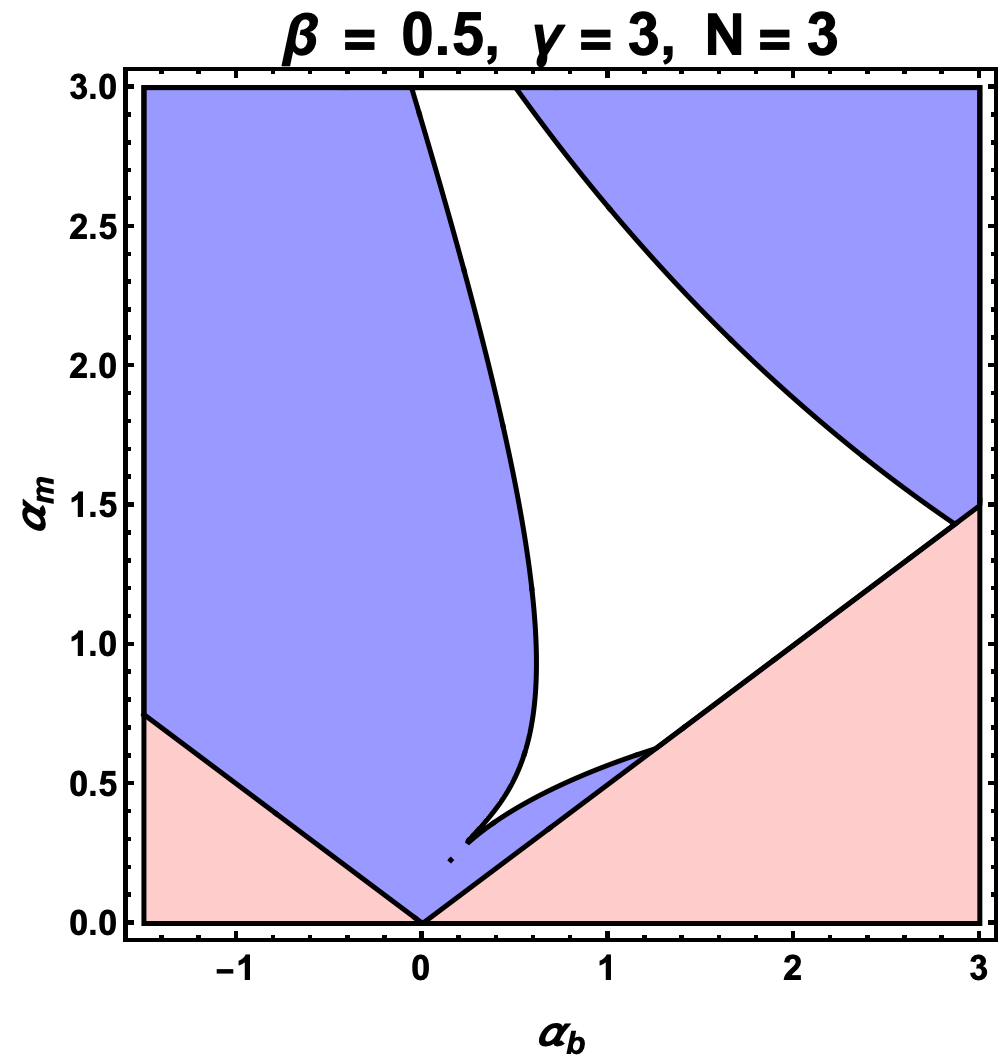}
    \caption{}\label{6-b}
  \end{subfigure}\hfill
  \begin{subfigure}[b]{0.32\textwidth}
    \centering
    \includegraphics[width=\textwidth]{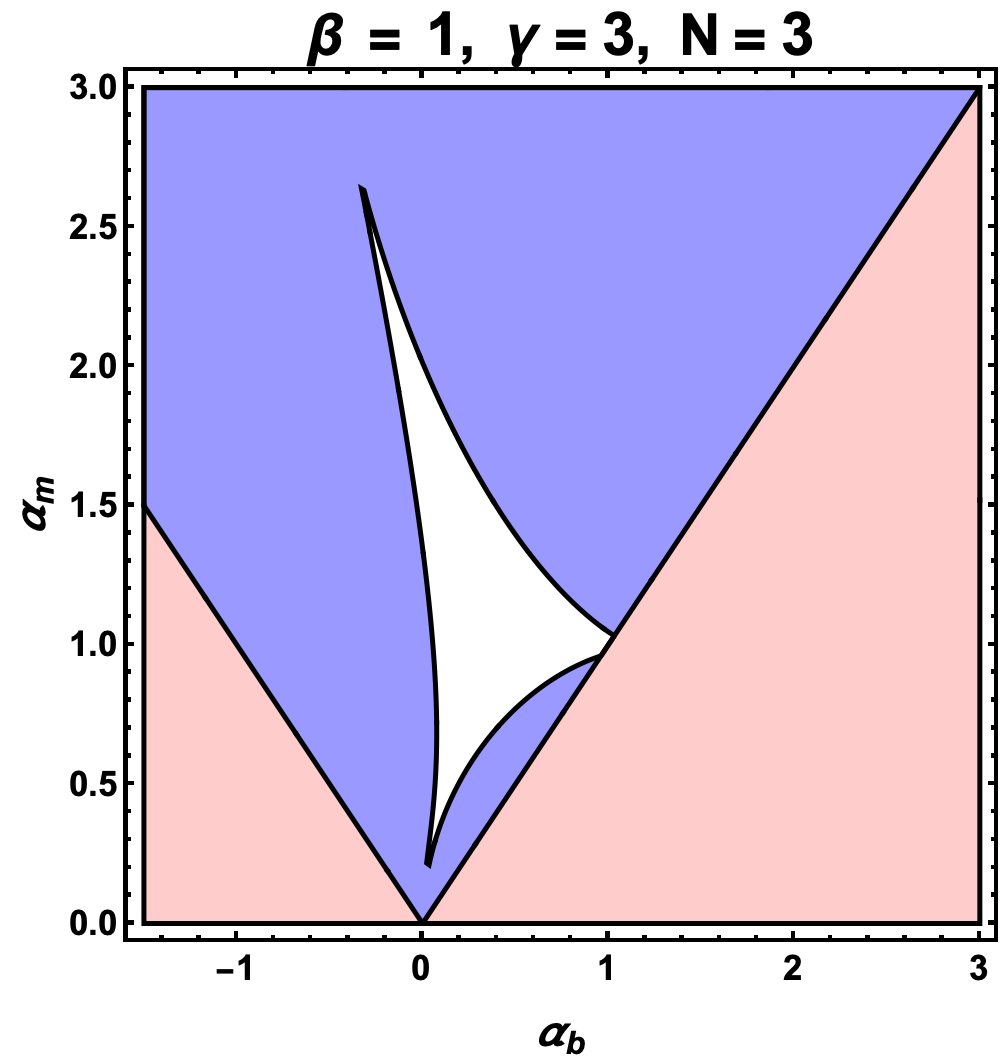}
    \caption{}\label{6-c}
  \end{subfigure}

  \vspace{0.5cm} 

  \begin{subfigure}[b]{0.32\textwidth}
    \centering
    \includegraphics[width=\textwidth]{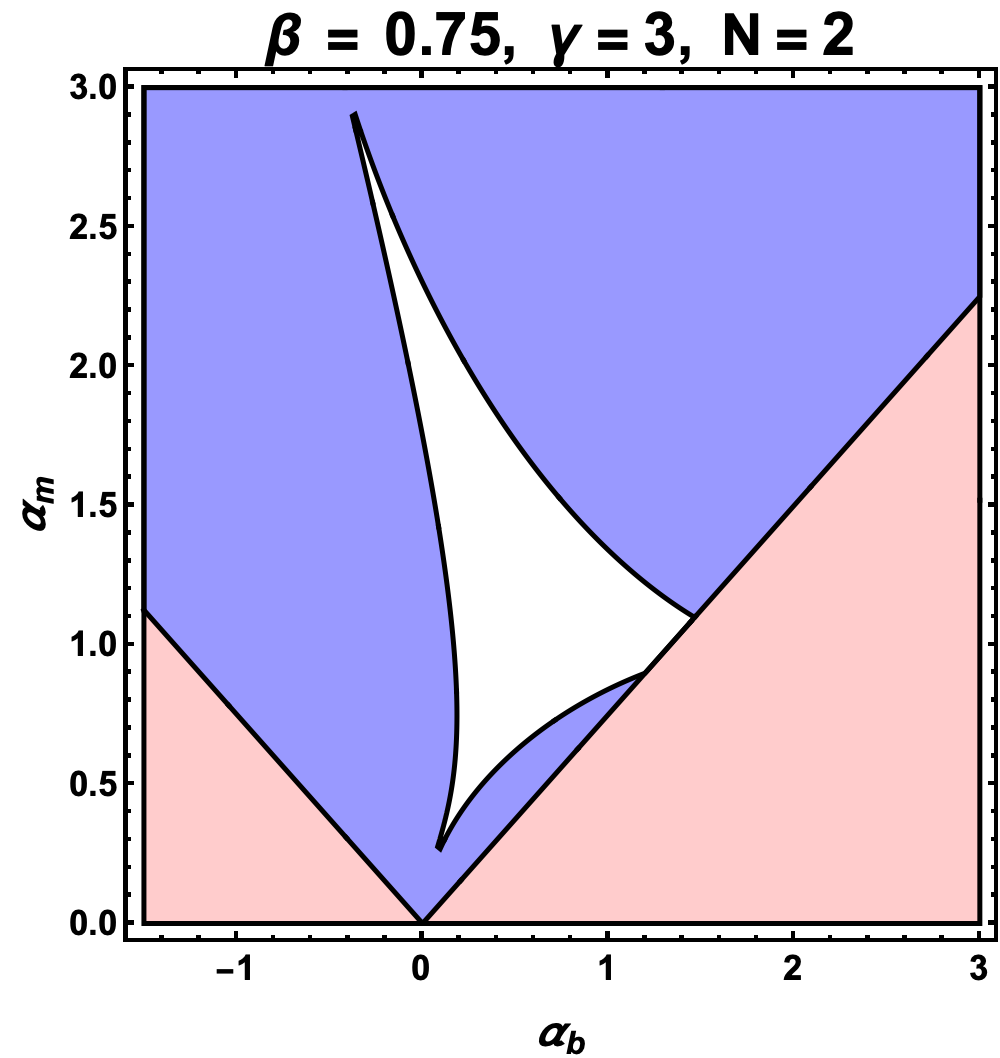}
    \caption{}\label{6-d}
  \end{subfigure}\hfill
  \begin{subfigure}[b]{0.32\textwidth}
    \centering
    \includegraphics[width=\textwidth]{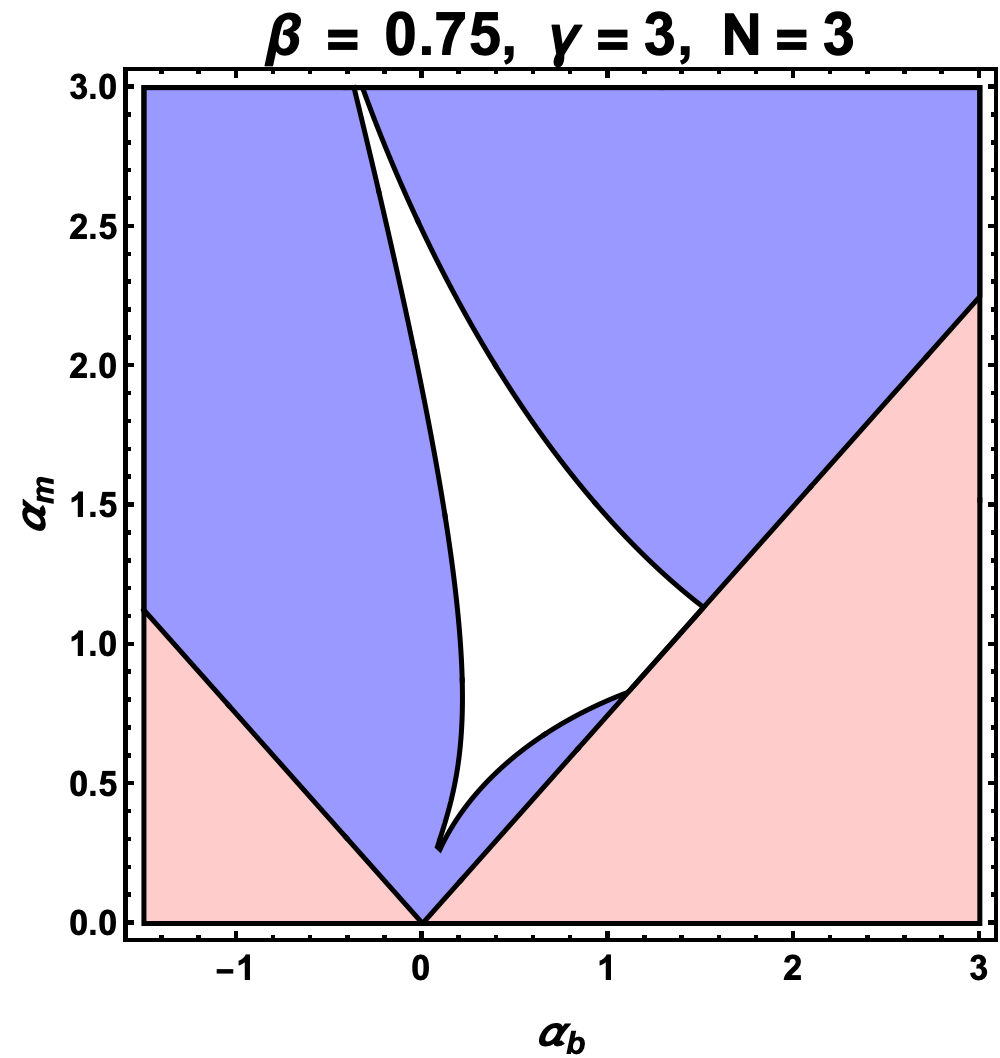}
    \caption{}\label{6-e}
  \end{subfigure}\hfill
  \begin{subfigure}[b]{0.32\textwidth}
    \centering
    \includegraphics[width=\textwidth]{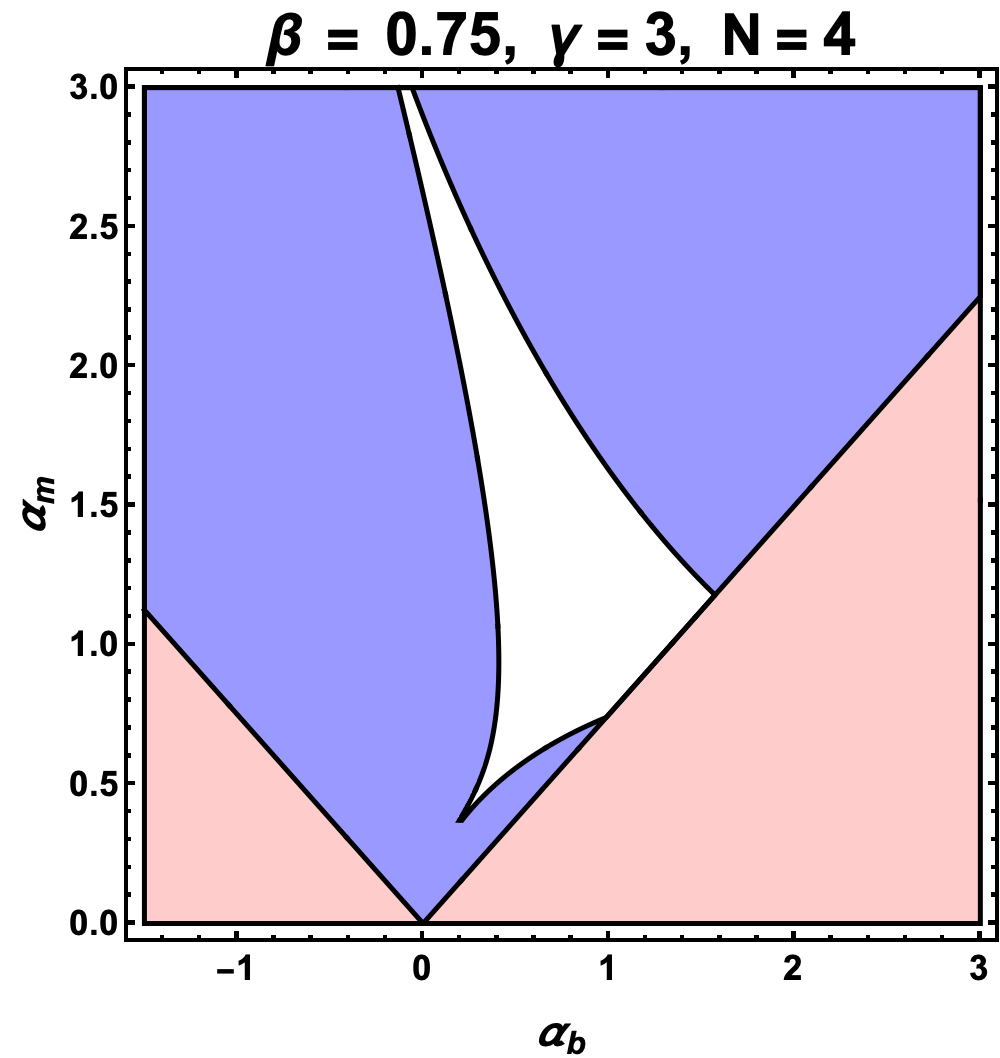}
    \caption{}\label{6-f}
  \end{subfigure}

  \caption{
  The parameter space of $\alpha_b$ and $\alpha_m$ is shown for $\gamma = 3$, with $\beta$ varying from $0.25$ to $1$ and $N = 2, 3$ and $4$. The couplings $\kappa_1$ and $\kappa_2$ are fixed to $1$.
  The color code is the same as that of Fig.~\ref{fig:β=0.75,Ν=3}}
  \label{fig:γ=3}
\end{figure}

Let us demonstrate the existence of a viable parameter space in which the total scalar potential $V_{\rm tot}$ remains bounded from below and $\theta$ is stabilized at a non-trivial value $(\neq0,\pi)$.
\FIG{fig:β=0.75,Ν=3} and \FIG{fig:γ=3} show the viable (white) regions in the $\alpha_b$-$\alpha_m$ plane for various parameter sets $(\beta,\gamma,N)$.
In the pink shaded regions, the potential is unbounded from below, as excluded by \EQ{bounded_cond}, while in the blue shaded regions, the potential is bounded from below but SCPV is not obtained. 
The white regions represent the viable parameter space which realizes CP-violating vacua.
More precisely, the white regions correspond to configurations in which both $v_1$ and $\theta$ are stabilized at $v_1 > 0$ and $-1 \leq \cos(2\theta) < 1$.
In contrast, although the potential remains bounded from below in the blue regions, the corresponding parameters do not produce a stationary point at $(v_1, \theta)$ satisfying $v_1 > 0$ and $-1 \leq \cos(2\theta) < 1$ with a positive-definite Hessian (mass-squared) matrix.

In \FIG{fig:β=0.75,Ν=3}, we fix $\beta = 0.75$ and $N = 3$, and vary $\gamma$. 
In panels \ref{5-a}–\ref{5-c}, the red curves denote the loci where the maximal determinant of the Hessian matrix, evaluated at the stationary points, vanishes.
For a given parameter set, the scalar potential may possess multiple stationary points obtained by solving the vanishing first derivatives with respect to $v_1$ and $\theta$.
Their corresponding Hessian matrices generally differ.
If, for all stationary points, the determinants of the Hessian matrices are not positive, the potential has no stabilized configuration.
Hence, the red curves delineate the critical boundaries in parameter space where the potential first develops a stable stationary point, defined by
$\max[\det(\text{Hessian})] = 0$.
When drawing the red curves, the physical constraint $-1 \leq \cos(2\theta) < 1$ is not imposed.
Instead, $\cos(2\theta)$ is replaced by a continuous parameter $y \in (-\infty, \infty)$, so that the red curves represent parameter sets yielding stationary points $(v_1, y)$ with $v_1 > 0$ and arbitrary real $y$ satisfying $\max[\det(\text{Hessian})] = 0$.
In panel \ref{5-a}, the left boundary of the white region coincides with the red curve, indicating that for small $\gamma$, the boundary of the viable parameter space is determined by the degeneracy of the stabilized vacua.
As $\gamma$ increases from $1$ to $3$, shown in panels \ref{5-b} and \ref{5-c}, the red curve gradually departs from the left boundary of the white region.
Numerical analysis shows that with increasing $\gamma$, this boundary transitions from being governed by $\max[\det(\text{Hessian})] = 0$ to being determined by the condition $\cos(2\theta) = 1$ at the stabilized configuration.

Meanwhile, the relatively small white regions in the panels \ref{5-d} and \ref{5-e} indicate that to make this approach plausible, $\gamma$ should not be significantly smaller or greater than $\mathcal{O}(1)$.
Since only the SUSY breaking terms cannot stabilize the flat direction as we mentioned in \SEC{sec:SUSYbreaking}, the dynamically generated potential term needs to be comparable to them.
In other words, we require $\gamma = \mathcal{O}(1)$, or
$m_{\rm soft}^2 \sim m_{\rm CP}^2 \lmk {\Lambda}/{m_{\rm CP}}\rmk^{6-2/N}$.
Thus the dynamical scale before the decoupling must be marginally smaller than the CP breaking scale, because of the hierarchy $m_{\rm soft}\ll m_{\rm CP}$.
This coincidence might be addressed by a more detailed model-building potentially with dynamical SUSY breaking, which is beyond the scope of our present paper and left for a future study.

In \FIG{fig:γ=3}, the panels \ref{6-a}, \ref{6-b}, \ref{6-e}, and \ref{6-c} correspond to cases where $\gamma = N = 3$ are fixed, and $\beta$ is varied. 
The boundary of the blue region is determined by $|\beta||\alpha_b| = \alpha_m$, with $0<|\beta| \leq 1$, as evident from Eq.~(\ref{dimlessV}) and consistent with Eq.~(\ref{bounded_cond}). 
Consequently, as $|\beta|$ increases while the other parameters are fixed, the areas of both the blue and white regions decrease.
Furthermore, as seen in the panels \ref{6-d}, \ref{6-e}, and \ref{6-f}, the viable parameter space exhibits little variation with changing $N$. 
This behavior arises because $N$ appears only in the exponent of the last term in Eq.~(\ref{dimlessV}), which is positive definite and bounded from below; 
its effect is further suppressed by a factor of $1/N$.

Let us present the masses of the modes $s(x)$ and $a(x)$. 
Since they are stabilized by soft SUSY breaking and non-perturbative effects, their masses are at the order of $\mathcal{O}(m_{\rm soft}^2)$.
By using \EQ{soft} and \EQ{Vdyn}, we obtain
\begin{align}
    m_a^2 &= \frac{\partial^2 V_{\rm soft}}{\partial a^2}+\frac{\partial^2 V_{\rm dyn}}{\partial a^2}\bigg|_{a, s \xrightarrow{} 0}
     =\frac{-2B^{(4)}_v}{f_a^2}\cos(2\theta)+\frac{N_a^{(4)}(\theta)}{f_a^2} \left(\frac{\Lambda^{2}}{D^{(2)}(\theta)}\right)^{3-1/N} ,   \label{mass_a} \\[3ex] 
    m_s^2 &= \frac{\partial^2 V_{\rm soft}}{\partial s^2}+\frac{\partial^2 V_{\rm dyn}}{\partial s^2}\bigg|_{a, s \xrightarrow{} 0}
    =\frac{B_v^{(4)}}{f_a^2} \cos(2\theta) + \frac{M_v^{(4)}}{f_a^2}+\frac{N_s^{(4)}(\theta)}{f_a^2}\left(\frac{\Lambda^2}{D^{(2)}(\theta)}\right)^{3-1/N},
     \label{mass_s}
\end{align}
in which we define
\begin{align} \label{mass_def}
    &B^{(4)}_v = b_1v_1^2+b_2v_2^2 \ ,\ \  \ \  M^{(4)}_v = m_1^2v_1^2+m_2^2v_2^2 \ , \notag \\[1ex]
    &N^{(4)}_a(\theta) = 4\left(1-\frac{1}{N}\right)\kappa_1 \kappa_2(\kappa_1^2+\kappa_2^2)[(\kappa_1^2v_1^2+\kappa_2^2v_2^2)\cos(2\theta)+2\kappa_1 \kappa_2v^2 \cos(4\theta)]v^2 \ , \notag \\[1ex]
    &N_s^{(4)}(\theta)=2\left(1-\frac{1}{N}\right)(\kappa_1^2+\kappa_2^2)\bigg[\left( 3-\frac{2}{N} \right) (\kappa_1^4v_1^4+\kappa_2^4v_2^4)+2\left(\frac{2}{N}-4+\cos(4\theta)\right)\kappa_1^2 \kappa_2^2 v^4\bigg] \ , \notag \\[1ex]
    &D^{(2)}(\theta) = (\kappa_1^2v_1^2+\kappa_2^2 v_2^2+2\kappa_1 \kappa_2 v^2 \cos(2\theta)) \ ,
\end{align}
and $f_a^2=v_1^2+v_2^2$ as we mentioned. 
Here, we find $B_v^{(4)} \sim M^{(4)}_v\sim v^2m_{1}^2$, $N_a^{(4)}(\theta) \sim N_s^{(4)}(\theta) \sim v^4$, and $D^{(2)}(\theta) \sim f_a^2 \sim v^2$. Through \FIG{fig:β=0.75,Ν=3}, we have concluded $\gamma = \mathcal{O}(1)$, which indicates that $\Lambda^{6-2/N} \sim m_1^2v^{4-2/N}$.
Substituting these back into Eq.~(\ref{mass_a}) and Eq.~(\ref{mass_s}), it turns out that $m_a^2 \sim m_s^2\lesssim m_{\text{soft}}^2 $.

\subsection{The Nelson-Barr framework
\label{sec:CKM}}

We now extend our setup to contain the SUSY Nelson-Barr mechanism to transfer SCPV into the SM sector \cite{Barr:1984qx,Nelson:1983zb,Barr:1984fh}.
Including additional quark multiplets $q,\bar{q}$, we consider the superpotential,
\beq
W_{\rm NB} = \mu\bar{q}q + (y_{1i} \Phi_1 + y_{2i} \Phi_2) q\bar{d}_i + Y_{Dij}H_dQ_{Li}\bar{d}_j \ ,
\eeq
where $\mu$ denotes the real mass parameter, $y_{\alpha i}$ is the real Yukawa coupling constant with $i,j=1,2,3$ and $\alpha=1,2$, and the last term is the SM Yukawa coupling for the down-type quarks ($Q_{Li}$ represent the left-handed SM quarks).
The charge assignments are shown in Table \ref{tab:charge}.
As is well-known, when $\phi_{1,2}$ obtain complex VEVs, the CKM phase is generated without reintroducing the strong CP phase.
Potentially dangerous terms which can spoil the mechanism are forbidden by a $Z_2$ symmetry.
Interestingly, the same $Z_2$ symmetry also forbids the soft breaking $A$-terms which would make the potential unbounded from below, as mentioned in \SEC{sec:SUSYbreaking}. 
Since one can see from \EQ{axion} that $\phi_1\propto e^{ia/f_a}$ and $\phi_2\propto e^{-ia/f_a}$, the model is nothing but the minimal Nelson-Barr (BBP) model \cite{Bento:1991ez}.

\begin{table*}[t]
\vspace{0mm}
\centering
\begin{tabular}{|c|c|c|c|c|c|c|c|c|c|}
\hline
& $\Phi_1$ & $\Phi_2$ & $X$ & $Q$ & $\bar{Q}$ & $q$ & $\bar{q}$ & $Q_{Li}$ & $\bar{d}_i$  \\
\hline
$SU(N)$ & ${\bm 1}$ & ${\bm 1}$ & ${\bm 1}$ & $\square$ & $\bar{\square}$ & ${\bm 1}$ & ${\bm 1}$ & ${\bm 1}$ & ${\bm 1}$ \\
$SU(3)_C$ & ${\bm 1}$ & ${\bm 1}$ & ${\bm 1}$ & ${\bm 1}$ & ${\bm 1}$ & $\square$ & $\bar{\square}$ & $\square$ & $\bar{\square}$\\
$U(1)_Y$ & $0$ & $0$ & $0$ & $0$ & $0$ & $\frac{1}{3}$ & $-\frac{1}{3}$ & $\frac{1}{6}$ & $-\frac{1}{3}$ \\
$Z_2$ & $-1$ & $-1$ & $1$ & $1$ & $-1$ & $-1$ & $-1$ & $1$ & $1$ \\
\hline
\end{tabular}
\vspace{3mm}
\caption{The charge assignments.}
\label{tab:charge}
\end{table*}

Let us briefly comment on an additional contribution to the scalar potential in this extension.
The coupling terms, $y_{\alpha i}\Phi_\alpha q\bar{d}_i$, break the spurious $U(1)$ symmetry, which can induce a potential term in the phase direction.
Such a potential could be generated by loop effects \cite{Davidi:2017gir} (see also Ref.~\cite{Dine:2024bxv}), but it is canceled in the exact SUSY limit.
Since the contribution is estimated as the order of the soft mass scale suppressed by loop factors, we expect that it is sub-dominant over the other terms and does not affect our conclusion.

\section{Conclusions and discussions}
\label{conclusion}

In this paper, we have explored the realization of SCPV in two distinct SUSY scenarios.
First, we investigated SCPV by SUSY-conserving dynamics, extending the spurion formalism developed in non-SUSY theories to analyze the stabilization of CP-violating phases, and incorporating an R-charge analysis to examine the stabilization of radial vacuum expectation values. Together, these analyses provide a systematic method to determine whether a given superpotential satisfies the necessary conditions for SCPV in exact supersymmetric vacua.
We showed a systematic procedure to judge the occurrence of SCPV in a general supersymmetric theory
(see Appendix \ref{sec:algorithm} for our program code).  
Second, we constructed a concrete model in which CP is spontaneously broken at an intermediate scale along pseudo-flat directions, stabilized by soft SUSY breaking and non-perturbative effects of a gauge theory. 
Our setup is consistently connected to the Nelson-Barr mechanism for inducing the CKM phase,
and predicts light scalars in the SCPV sector whose masses are determined by the soft mass scale.

In the model of \SEC{susy-breaking case}, the scalar fields, $a$ and $s$, acquire masses $\sim m_{\rm soft}$.
The (axino-like) fermion partners also obtain a mass due to the SUSY breaking effects.
Although it is highly model-dependent, for example, a Planck-suppressed operator with SUSY breaking fields can give rise to the mass of the order of the gravitino mass $m_{3/2}$.
Since gauge-mediated SUSY breaking (see e.g. Refs.~\cite{Giudice:1998bp,Kitano:2010fa} for reviews) is considered as a plausible possibility to suppress dangerous corrections to $\bar{\theta}$ \cite{Dine:1993qm,Dine:2015jga}, the axino-like particle and gravitino masses are smaller than the soft mass scale.
Both are thermally produced and behave as cold dark matter, which typically leads to a severe upper bound on the reheating temperature $T_{\rm reh}$ \cite{Cheung:2011mg}.
In addition, the scalar components $a, s$ can be thermally/non-thermally produced.
In our setup (or the BBP model), the CP breaking fields are coupled to the down-type quarks and the Nelson-Barr heavy fermion, and the decay of the scalar fields may be suppressed for a large $m_{\rm CP}$.
In the present paper, we have focused on the stabilization mechanism of CP breaking fields, but its phenomenological and cosmological consequences should be interesting and is left for a future exploration.

If CP symmetry is spontaneously broken after inflation, there appear domain walls associated with the CP breaking. 
The energy density of stable domain walls easily dominates that of the Universe, and the standard cosmology would be spoiled.
In the case of gauged CP which is motivated from the viewpoint of the CP quality, the situation would be worse \cite{McNamara:2022lrw,Asadi:2022vys}. 
To avoid the disastrous situation, we need to require that CP symmetry is spontaneously broken before inflation or not restored after inflation. 
For the coupling constants of the order of unity, the maximum temperature in the Universe should be lower than the CP breaking scale $m_{\rm CP}$ in both scenarios of SCPV stabilized (non-)supersymmetrically.
As an alternative prescription, we can consider some scalar fields coupled to the CP breaking fields with negative coupling coefficients.
The induced thermal potential can interrupt the restoration of CP symmetry after inflation, which relaxes the bound on $m_{\rm CP}$ significantly \cite{Dvali:1995cc,Dvali:1996zr}.\footnote{This approach has been considered as a solution to the axion domain wall problem, and can be similarly applied to the CP domain wall, as mentioned in these references.
Recently, the thermal non-restoration of the Peccei-Quinn symmetry was discussed in Ref.~\cite{Nakagawa:2025suc} for a model similar to the present model.}

\acknowledgments

We would like to thank Jason Evans for useful discussions.
YN is supported by Natural Science Foundation of Shanghai.

\appendix

\section{The N chiral superfields
\label{sec:spurion}}

We consider a renormalizable superpotential of $N$ superfields $\Phi_i$ ($i = 1,2, \cdots, N$),
\begin{equation}
  W = L^i \Phi_i + \frac{1}{2} M^{ij}\Phi_i \Phi_j +\frac{1}{6}  Y^{ijk} \Phi_i \Phi_j \Phi_k \ ,
  \label{renormalizable}
\end{equation}
where $M^{ij}$ is symmetric under the exchange of indices $i$ and $j$, and $Y^{ijk}$ is totally symmetric under any permutation of $i$, $j$, and $k$. 
Here we focus on the case that any of the coefficients are not restricted e.g. by symmetry.
The $F$-term potential is given by
\begin{align}\label{2-2}
V(\phi_i, \phi^{*i})
&=\sum_i \left|\frac{\del W}{\del\phi_i}\right|^2 \notag \\
&= L^*_i L^i + M^*_{ij}M^{ik} \phi^{*j} \phi_k + \frac{1}{4}Y^*_{ijk}Y^{ilm}\phi^{*j} \phi^{*k} \phi_{l} \phi_{m} \notag \\
&+ \left( L^*_i M^{ij} \phi_j + \frac{1}{2} L^*_i Y^{ijk} \phi_j \phi_k + \frac{1}{2} M^*_{ij}Y^{ikl} \phi^{*j}\phi_k \phi_l + {\rm h.c.} \right).
\end{align}
By imposing exact CP symmetry on the Lagrangian, we take a real basis in which all the coefficients $\{L^i, M^{ij}, Y^{ijk}\}$ in the superpotential are real without loss of generality.\footnote{
Since superpotential is a holomorphic function, it is the Lagrangian that is invariant under CP transformation. 
Combined with the symmetry properties of $M^{ij}$ and $Y^{ijk}$, it turns out that the phases of all the coupling coefficients must be identical up to a factor of $\pi$, 
and thus, we can set them to zero. 
} 

\renewcommand{\arraystretch}{1.5} 

\begin{table}[!t] 
\begin{tabular}{|c|c|c|c|}
  \hline
  Spurion & Charges & Condition & Number \\ 
  \hline
   $M^*_{ij}M^{ik}$& $\pm (..., \underset{j}{+1}, ..., \underset{k}{-1},...)$& $j \neq k, \, N\geq 2$ & $C^2_N$ $\ast$ \\
  \hline
  $Y^*_{ijk} Y^{ilm}$ &   
  \begin{tabular}{@{}c@{}}
   $\pm (..., \underset{k}{+1}, ..., \underset{m}{-1},...)$ \\
  \hline
  $\pm (..., \underset{j=k}{+2}, ..., \underset{l=m}{-2},...)$ \\
  \hline
    $\pm (..., \underset{j=k}{+2}, ..., \underset{l}{-1},..., \underset{m}{-1},...)$ \\
   \hline
     $\pm (..., \underset{j}{+1}, ..., \underset{k}{+1},...,\underset{l}{-1},..., \underset{m}{-1},...)$\end{tabular}&   
  \begin{tabular}{@{}c@{}}
    $j = l \, , k \neq m, \, N \geq 2$ \\
   \hline
    $j = k \neq l = m, \, N \geq 2$ \\
  \hline
    $j = k \neq l \neq m, \, N \geq 3$ \\
    \hline
    $ j \neq k \neq l \neq m , \, N \geq 4$ \\
  \end{tabular} &   \begin{tabular}{@{}c@{}}
    $C^2_N$ $\ast$ \\
    \hline
    $C^2_N$ \\
    \hline
    $C^1_N C^2_{N-1}$ \\
    \hline
    $\frac{1}{2}C^2_N C^2_{N-2}$ \\
  \end{tabular} \\ 
  \hline 
  $L^*_i M^{ij}$ & $\pm (..., \underset{j}{+1}, ...)$& $N \geq 1$ & $N$ $\star$\\
  \hline
  $L^*_i Y^{ijk}$ &   \begin{tabular}{@{}c@{}} 
    $\pm (..., \underset{j=k}{+2},...)$ \\
    \hline
    $\pm (..., \underset{j}{+1},..., \underset{k}{+1},...)$ 
  \end{tabular}&   
  \begin{tabular}{@{}c@{}} 
    $j = k, \, N \geq 1$ \\
    \hline
    $j \neq k,\, N \geq 2$ \\ 
  \end{tabular} &   \begin{tabular}{@{}c@{}}
    $N$ \\
    \hline
    $C^2_N$ \\
  \end{tabular}   \\
  \hline
  $M^*_{ij} Y^{ikl}$ &   \begin{tabular}{@{}c@{}} 
    $\pm(..., \underset{l}{+1}, ...) $ \\
    \hline
     $\pm(..., \underset{j}{+1}, ..., \underset{k=l}{-2}, ...  )$ \\
    \hline
     $\pm(..., \underset{j}{+1}, ..., \underset{k}{-1}, ..., \underset{l}{-1},...  )$
  \end{tabular}&   \begin{tabular}{@{}c@{}}
    $ j=k, N \geq 1$ \\
    \hline
    $j \neq k = l, \, N \geq 2$ \\
    \hline
    $j \neq k \neq l,  \, N \geq 3 $ \\
  \end{tabular}  &   \begin{tabular}{@{}c@{}} 
    $N$ $\star$ \\
    \hline
    $C^1_N C^1_{N-1}$ \\
    \hline
    $C^1_N C^2_{N-1}$ \\
  \end{tabular}  \\
  \hline
\end{tabular}
\vspace{3mm}
\caption{
The summary of spurions in the scalar potential and their corresponding charges.
The $\ast$ and $\star$ represent the different pairs of multiplicity, which refer to cases where different spurions share equivalent charge (i.e., their charges differ only by an overall sign).
}
\label{Tab_1}
\end{table}

In such an $N$-superfield theory, a subgroup of the maximal symmetry group satisfied by the kinetic terms can be taken as $U(1)_1 \times U(1)_2 \times \cdots \times U(1)_N \equiv \otimes_N U(1)$.
Each $U(1)_k$ rotates the phase of one complex degree of freedom, such that
\beq
&& \Phi_j \xrightarrow{U(1)_k} e^{i {\beta}_k \hat{X}_k} \Phi_j
= e^{i {\beta}_k \delta_{kj}} \Phi_j \ ,
\label{U1tr1}\\[1ex]
&& \Phi^{*j} \xrightarrow{U(1)_k} e^{-i {\beta}_k \delta_{kj}} \Phi^{*j} \ ,
\label{U1tr2}
\eeq
where $\beta_k$ and $\hat{X}_k$, respectively, represent a real constant and the generator for $U(1)_k$.
Note that the sum of $k$ is not taken.

Let us promote the coefficients in the superpotential to the spurions with charges under $U(1)_k$ symmetries, so that the explicit breaking of $\otimes_N U(1)$ is formally restored.
We write down the transformation law for the couplings in the superpotential:
\beq
&& L^i \xrightarrow{U(1)_k} e^{-i \beta_k \delta_{ki}} L^i \ ,\\[1ex]
\label{2-5}
&& M^{ij} \xrightarrow{U(1)_k} e^{-i {\beta}_k (\delta_{ki}+\delta_{kj})} M^{ij} \ ,\\[1ex]
\label{2-6}
&& Y^{ijl} \xrightarrow{U(1)_k} e^{-i {\beta}_k (\delta_{ki}+\delta_{kj}+\delta_{kl})} Y^{ijl} \ .
\label{2-7}
\eeq
We then define a charge for each spurion as an $1 \times N$ row vector whose $k$-th component is filled with the charge of the spurion under $U(1)_k$. 
For instance, the charge vector for $L^i$ is $\bm{Q}^{\ (L^i)} = (0,...,0,\underset{i}{-1},0,...,0)$. 
In the following context, for simplicity and cleanness, we only write down the non-zero components, like $\bm{Q}^{\ (L^i)} = (\cdots,\underset{i}{-1},\cdots)$. 
Since the scalar potential is also invariant under $\otimes_N U(1)$, the coefficients which are products of spurions in the superpotential are identified as new spurions.
Note that for any spurion in the scalar potential with a $U(1)_k$ charge $\bm Q_k$, there exists a complex conjugate spurion with a $U(1)_k$ charge $-\bm Q_k$ due to the hermitian nature of the Lagrangian. 
Therefore, it is the magnitude of the charge that is physically significant.
If the charges of two spurions are identical up to an overall minus sign, we consider these two spurions to be equivalent.
Inequivalent spurions actually induce different forms of potential terms on phases, and a physical CP phase can be generated, as long as the number of independent spurions is large enough.

We systematically identify and summarize all possible spurions and their corresponding charges for the scalar potential in Table~\ref{Tab_1}.
The third column shows the condition for the existence of the corresponding charge vectors, while the fourth one does the number of independent spurions.
We note that some spurions of different forms share equivalent charges, indicated by $\ast$ and $\star$ in Table \ref{Tab_1}, i.e. they are not independent spurions.
One can also see that the number of possible charge vectors or independent spurions depends on $N$.
For example, the charge vector $\bm{Q}^{\, (Y^*_{ijk} Y^{ilm})} = \pm (..., \underset{j}{+1}, ..., \underset{k}{+1},...,\underset{l}{-1},..., \underset{m}{-1},...)$ requires all four indices $(j,k,l,m)$ to take a different number, and thus, $N\geq4$.

\section{Program code for judging the existence of physical CP violation
\label{sec:algorithm}}

\lstset{
  language=Mathematica,
  basicstyle=\ttfamily\small,
  keywordstyle=\color{blue!70!black},
  commentstyle=\color{gray!80!black}\itshape,
  stringstyle=\color{orange!60!black},
  breaklines=true,
  columns=fullflexible,
  showstringspaces=false,
  frame=tb,
  rulecolor=\color{black!20},
  numbers=left,
  numberstyle=\tiny\color{gray},
  xleftmargin=2em,
  framexleftmargin=1.5em
}

\lstset{
  literate={μ}{{\textmu}}1
           {η}{{\ensuremath{\eta}}}1
           {Φ}{{\ensuremath{\Phi}}}1
           {φ}{{\ensuremath{\phi}}}1
           {ψ}{{\ensuremath{\psi}}}1
           {Π}{{\ensuremath{\Pi}}}1
           {Σ}{{\ensuremath{\Sigma}}}1
           {Λ}{{\ensuremath{\Lambda}}}1
           {Ω}{{\ensuremath{\Omega}}}1
           {κ}{{\ensuremath{\kappa}}}1
           {χ}{{\ensuremath{\chi}}}1
           {θ}{{\ensuremath{\theta}}}1
           {ρ}{{\ensuremath{\rho}}}1
           {β}{{\ensuremath{\beta}}}1
           {γ}{{\ensuremath{\gamma}}}1
           {Δ}{{\ensuremath{\Delta}}}1
           {α}{{\ensuremath{\alpha}}}1
           {λ}{{\ensuremath{\lambda}}}1
           {Ξ}{{\ensuremath{\Xi}}}1
           {ξ}{{\ensuremath{\xi}}}1
           {→}{{$\to$}}1
           {∞}{{$\infty$}}1
           {⟩}{{$\rangle$}}1
           {⟨}{{$\langle$}}1
}

\begin{lstlisting}[caption={Mathematica code for SUSY spurion analysis}, label={lst:spurion}]

(*//////////////////////////////////////////////////////////////*)
(* Sec. I: Define the state vector |W(Φ)⟩ = WΦ[nF_] *)
(*//////////////////////////////////////////////////////////////*)
Clear[Φ]
(*Define the notation of superfields*)
fields[nF_] := Array[Φ, nF];

(*------An Example-----*)
(*W=Xμ^2+X(aη_1^2+bη_1η_2+cη_2^2)+Y(dη_1^2+eη_1η_2+fη_2^2)*)
(*Φ[1]=η_1, Φ[2]=η_2, Φ[3]=X, Φ[4]=Y*)

WΦ[nF_] := Module[{φs = fields[nF]}, 
            μ^2*Φ[3] + (a*Φ[1]^2 + b*Φ[1]Φ[2]+cΦ[2]^2 )*Φ[3] 
            + (d*Φ[1]^2+eΦ[1]Φ[2]+fΦ[2]^2)*Φ[4]];
(*The number of chiral superfields*)
numF = 4;

(*//////////////////////////////////////////////////////////////*)
(* Sec. II: Define the bra ⟨F_Q(Φ^m)| and inner product *)
(*//////////////////////////////////////////////////////////////*)
(*Length of the charge vector*)  
NQ[Q_] := Length[Q];
(*Define the Bra vector <F_ Q(Φ^m)|*)
Bra[x_, n_] := Function[expr, D[expr / n!, {x, n}] /. {x -> 0}];
braF[Q_, expr_] := Fold[Bra[Φ[#2], Q[[#2]]][#1] &, expr, Range[Length[Q]]];
(*If considering non-renormalizable theory, change 3 to a larger number*)

(*//////////////////////////////////////////////////////////////*)
(* Sec. III: Extract the spurion s^Q *)
(*//////////////////////////////////////////////////////////////*)

(* ===Generate all possible charge vectors for nF superfields and Φ^m===*)
ChargeVectors[nF_, nPower_] := 
  Module[{Q}, 
   Q = Sort[
      Flatten[Permutations /@ IntegerPartitions[nPower + nF, {nF}], 
       1]] /. n_?NumberQ :> n - 1;
   {Length[Q], Q}];
   
s[Q_, nF_] := braF[Q, WΦ[nF]]; (*Define s[Q,nF] = ⟨F_Q(Φ^m)|W(Φ)⟩*)

(* ===List all s^Q for a SUSY theory with nF fields===*)
sListRaw[nF_] := 
  Flatten[Table[
    With[{chargeVectors = ChargeVectors[nF, nPower]}, 
     Table[{s[chargeVectors[[2, i]], nF], nPower, 
       chargeVectors[[2, i]]}, {i, chargeVectors[[1]]}]], {nPower, 1, 
     3}], 1];

(* ===Filter out vanishing coupling coefficients===*)
sList[nF_] := Select[sListRaw[nF], FreeQ[First[#], 0] &];
sList[numF]; 
(*The output is in the form {s^Q, m(power of the superfield basis), Q}*)

(*//////////////////////////////////////////////////////////////*)
(* Sec. IV: Construct the spurion of scalar potential {s~}_{Q'}^{Q} *)
(*//////////////////////////////////////////////////////////////*)

(*===List all the {s~}_{Q'}^{Q}===*)
qUnit[i_, nF_] := UnitVector[nF, i];
sTildeList[nF_] := 
  Module[{mPower, nPower, cVm, cVn, Qm, Qn}, 
   Flatten[Table[cVm = ChargeVectors[nF, mPower];
      cVn = ChargeVectors[nF, nPower];
      Table[Qm = cVm[[2, i]];
       Qn = cVn[[2, j]];
       If[mPower <= nPower,
        With[{sum = 
           Total[Table[
             s[Qm + qUnit[k, nF], nF]*
              SuperStar[s[Qn + qUnit[k, nF], nF]], {k, nF}]]}, {sum, 
          mPower, nPower, Qm, Qn, Qm - Qn}], Nothing], {i, 
        cVm[[1]]}, {j, cVn[[1]]}], {mPower, 0, 2}, {nPower, 0, 2}], 3
        ] /. SuperStar[0] -> 0];
(*If considering non-renormalizable theory, change 2 to a larger number*)

(*//////////////////////////////////////////////////////////////*)
(* Sec. V: Organize the output results *)
(*//////////////////////////////////////////////////////////////*)

SimplifySpurions[data_] := 
  Module[{filtered, grouped, summed},
   (*Step 1: Discard entries with zero first element (spurion)*)
   filtered = Select[FullSimplify[data], FreeQ[First[#], 0] &];
   (*Step 2:Group by the last element (charge vector Q)*)
   grouped = GatherBy[filtered, Last];
   (*Step 3:Sum over first elements within each group*)
   summed = {Total[First /@ #], Last[First[#]]} & /@ grouped;
   (*Step 4:Remove pairs that differ only by a total minus sign*)
   Fold[Function[{acc, elem}, 
     If[MemberQ[acc[[All, 2]], -elem[[2]]], acc, 
      Append[acc, elem]]], {}, summed]];

SimplifySpurions[sTildeList[numF]]
(*The output is in the form of {(Σ_{Q'} {s~}_{Q'}^{Q}), Q} *)

(*//////////////////////////////////////////////////////////////*)
(* Sec. VI: Check the SCPV condition *)
(*//////////////////////////////////////////////////////////////*)
AnalyzeSpurions[results_] := 
  Module[{vectors, nonTrivial, mat, rank, 
    nRows},(*Extract charge vectors and remove trivial ones*)
   vectors = Last /@ results;
   nonTrivial = Select[vectors, ! VectorQ[#, # == 0 &] &];
   mat = nonTrivial;
   rank = MatrixRank[mat];
   nRows = Length[mat];
   Print["Non-trivial charge vectors:\n", nonTrivial];
   Print["Matrix form:\n", mat];
   Print["Rank of matrix: ", rank];
   Print["Number of inequivalent spurions: ", nRows];
   If[nRows > rank, 
    Print["Yes! This model satisfies the necessary condition of SCPV."],
     Print["NO! This model DOES NOT satisfy the necessary condition of SCPV."]];
   <|"NonTrivialVectors" -> nonTrivial, "Matrix" -> mat, 
    "Rank" -> rank, "NumRows" -> nRows|>];

AnalyzeSpurions[SimplifySpurions[sTildeList[numF]]];
\end{lstlisting}

\bibliography{reference}

\end{document}